%% file: RidgeShrinkage.tex
\documentclass[a4paper,11pt,times,authoryears]{article}
\usepackage{authblk}
\usepackage[authoryear]{natbib}
\usepackage{bm}
\usepackage{float}
\usepackage{amsmath}
\usepackage{amsthm}
\usepackage{mathtools}
\usepackage{amssymb,amsfonts}
\usepackage{graphicx}
\usepackage{mathrsfs}
\usepackage{multirow,multicol}
\usepackage{enumitem, array}
\usepackage{subcaption}
\usepackage{epstopdf}
\usepackage{float}
\usepackage{enumerate}
\usepackage{booktabs}
\usepackage{setspace}
\usepackage{adjustbox}
\usepackage{pdflscape}
\usepackage[colorlinks,citecolor=blue]{hyperref}
\usepackage{rotating}
\usepackage{actuarialsymbol}
\usepackage{fullpage}
\usepackage{setspace}
\usepackage{lscape}
\usepackage{booktabs}
\usepackage{setspace}
\usepackage{tocbibind}
\usepackage{color}
\usepackage{commath} 
\usepackage{setspace}
\usepackage{algorithm}
\usepackage{rotating}
\usepackage[noend]{algpseudocode}

\newtheorem{Theorem}{Theorem}[section]

\newtheorem{Lemma}[Theorem]{Lemma}

\usepackage{geometry}
\def\bC{{\mathbf C}}
\def\bD{{\mathbf D}}

\def\bH{{\mathbf H}} 
\def\bI{{\mathbf I}} 
\def\bK{{\mathbf K}}

\def\bT{{\mathbf T}}

\def\bW{{\mathbf W}}

\def\bX{{\mathbf X}}

\def\bc{{\mathbf c}}

\def\bx{{\mathbf x}}

\def\by{{\mathbf y}}

\def\by{{\mathbf y}}

\def\bbeta{{\boldsymbol{\beta}}}
\def\bmu{{\boldsymbol{\mu}}}

\def\hbbeta{\widehat{\boldsymbol \beta}}

\def\Wh{\widehat{\bW}}

\def\bzero{{\mathbf 0}}

\def\mle{\widehat\bbeta}
\def\rmle{\widehat\bbeta_{\textrm{RMLE}}}
\def\ur{\widehat{\boldsymbol{\beta}}^{UR}}
\def\rr{\widehat{\boldsymbol{\beta}}^{RR}}
\def\rls{\widehat{\boldsymbol{\beta}}^{RLS}}
\def\rpt{\widehat{\boldsymbol{\beta}}^{RPT}}
\def\spe{\widehat{\boldsymbol{\beta}}^{SPE}}
\def\rs{\widehat{\boldsymbol{\beta}}^{RS}}
\def\rps{\widehat{\boldsymbol{\beta}}^{RPS}}
\begin{document}
\begin{small}
\title{Ridge--Type Shrinkage Estimators in  Low and High Dimensional Beta Regression Model with   Application
in Econometrics and Medicine}
\author{ \ Ejaz Ahmed$^1$, \  Reza Arabi Belaghi$^2$, \ Yasin Asar$^3$,  \ Abdulkhadir Hussein$^2$ \footnote{Corresponding author: Email: ahussein@uwindsor.ca}
 \\
  \emph{\small{$^1$Department of
Mathematics and Statistics, Brock University, St. Catharines, ON, Canada\\}}
 \emph{\small{$^2$Department of Mathematics and Statistics, University of Windsor, Windsor, ON, Canada\\}}
 \emph{\small{$^3$Department of Mathematics and Computer Sciences, Necmettin Erbakan University, Konya, Turkey}}

}
\date{}
\maketitle
\begin{abstract}
Beta regression (BR) model is useful in the analysis of bounded continuous outcomes  such as proportions. It is well-known that for any regression model, the presence of multicollinearity leads to poor performance of the maximum likelihood (ML) estimators. The ridge--type estimators have been proposed to alleviate the adverse effects of the multicollinearity. Furthermore, when  some predictors have insignificant or weak effects on the outcomes, it is desired to recover as much information as possible from these predictors instead of discarding them all together. In this paper, we proposed ridge--type shrinkage estimators for the low and high dimensional BR model, which address the above two issues simultaneously. We compute the biases and variances of the proposed estimators in closed forms and use Monte Carlo simulations to evaluate their performances. The results show that, both in low and high dimensional data,  the performance of the proposed estimators are superior to ridge estimators that discard weak or insignificant predictors. We conclude this paper by applying the proposed methods for two real data from econometric and medicine. 


\vskip 6mm
\noindent{Supplementary Information (SI): Appendix 1, Appendix 2, Appendix 3.}
\end{abstract}
\end{small}

\noindent
\section{\large{Introduction}}
\label{sec1}
In modeling the relationship between predictor variables  and a bounded continuous outcome variable,   the usual linear regression or  gamma regression models are not appropriate. Beta regression (BR) model is a good alternative in such situations. \citet{Ferrari} pioneered the application of the beta regression models for modeling the  proportions in the unit intervals along with some predictor variables. The BR   model has been applied in various scientific areas of research such as modelling the proportion of income
spent on food, the poverty rate, the proportion of crude oil converted to gasoline and the proportion of surface covered by vegetation \citep{Qasim}. The BR has also been applied in  modeling bounded time series data such as the of Canada's  Google® Flu Trends, \citep{Guolo} as well as in machine learning  \citep{Espinheira}.  

Usually, the maximum likelihood estimator (MLE) is used to estimate the unknown regression coefficients \citep{Ferrari}. Multicollinearity problem may arise when there are near linear dependencies among the  predictor variables. This problem leads to high variance of the estimated coefficient vectors, which in turn may jeopardize the interpretability of the coefficients as well as any statistical significance  testing procedure based on them. In order to alleviate the negative effects of multicollinearity, the Ridge estimators are often employed  \citep{Hoerl1970}. Recently, \citet{Qasim}   generalized   the usual ridge estimation  methodology to the case of the beta regression model. 

In regression models, when there are some prior information about parameter vector $ \boldsymbol{\beta} $ under a linear restriction such as $ \textbf{H} \boldsymbol{\beta} = \textbf{h} $, the shrinkage strategies, namely linear shrinkage \citep{Thompson1968}, pretest \citep{Bancroft1944}, shrinkage pretest estimator \citep{Ahmed1992}, Stein estimator \citep{Stein1956}, and positive Stein estimators are applied to the estimation of parameters. Based on these linear restrictions, the parameter vector $ \boldsymbol{\beta} $ is partitioned into two parts as $ \boldsymbol{\beta} = (\boldsymbol{\beta}_1^\prime, \boldsymbol{\beta}_2^\prime)^\prime $ where $ \boldsymbol{\beta}_1 $ contains the active or significant parameters and $ \boldsymbol{\beta}_2 $ contains the inactive parameters that are not significant in predicting the outcome. Therefore, two models are considered. One is a full model or unrestricted model and includes all parameters that are estimated with the maximum likelihood method. The other model is a sub--model or restricted model, which contains only the significant parameters. For more details about methodology of shrinkage estimations, we refer the reader to  \citet{Ahmed2014}. 

The primary motivation of this paper is to develop ridge--type shrinkage estimators for both low and high dimensional beta regression model in the presence of highly correlated predictor variables among which there are some  insignificant (or week) effect on the outcome of interest.  Such estimators are expected to be more efficient than the usual ridge estimators that only alleviate the problem of multicollinearity. We study the theoretical properties of the proposed methods and design a Monte Carlo simulation study to compare their relative performance with respect to the usual ridge--type unrestricted estimator.  

The paper is organized as follows: In \ref{sec2}, we introduce the beta regression model and the  ridge--type unrestricted estimators of the regression effects. We then  construct our proposed ridge--type shrinkage estimators. In Section \ref{sec3}, the asymptotic behavior of the bias   variance of the proposed estimators are presented. We conduct a Monte Carlo simulation study to compare the performance of the proposed estimators in Section \ref{sec4}. We apply the proposed estimation methods to two real data set in Section \ref{app}. Finally, conclusions are presented in Section \ref{conc}.


\section{Theory and Method}\label{sec2}
In this section, we introduce the beta regression model and define  the ridge--type unrestricted and restricted estimators. We then introduce the ridge--type shrinkage estimators of the parameters.

\subsection{Beta Regression Model}
Following \citet{Ferrari} and \citet{Qasim}, let $\by=\left[y_1, y_2, \ldots, y_n\right]^\prime$ be independent observations of the response variable following a (reparameterised) beta distribution with probability density function  
\begin{equation}\label{beta:pdf2}
    f( y_{i}; \mu , \phi)=\frac{\Gamma(\phi)}{\Gamma(\mu\phi)\Gamma(\phi(1-\mu))}y^{\mu\phi-1} (1-y)^{(1-\mu)\phi-1},	\ y \in (0,1),
\end{equation}
where $0<\mu<1$ and $\phi>0$ such that $y_i \sim Beta\left(\mu \phi, (1-\mu)\phi\right)$. Therefore, the mean and variance of this distribution are, respectively, $E(y_i)=\mu$ and $Var(y_i)=V(\mu)/(1+\phi)$ where $V(\mu)=\mu(1-\mu)$. 

Using a link function $g(.)$, one can define a  beta regression model by  
\begin{equation}\label{lin:pred}
    g(\mu_i)=\sum_{j=1}^p x_{ij}\beta_j=\bx_i^\prime\bbeta=\eta_i
\end{equation}
where $\bx_i^\prime$ is the ith   vector of covariates, and  $\bbeta=\left[\beta_1, \beta_2, \ldots, \beta_p \right]^\prime$ is a vector of regression parameters (effects). To simply the notations, we will lump all covariate vectors in a design matrix $\bX=\left[\bx_1^\prime,\bx_2^\prime,\ldots,\bx_n^\prime\right]$  of order $n \times p, (n>p)$. 

In Equation \ref{lin:pred}, we assume that the link function $g(.)$ is a strictly monotone and twice differentiable function defined   $(0,1)$ with values in   $\mathbb{R}^p$.

Although different link functions are available for beta regression model \citep{Ferrari}, we prefer to use the logit link function $g(\mu)=\log(\mu/(1-\mu)$ so that 
\begin{equation}\label{logit}
    \mu_i=\frac{\exp(\bx_i^\prime\bbeta)}{1+\exp(\bx_i^\prime\bbeta)}
\end{equation}
for $i=1,2, \ldots, n$.
Thus, the corresponding log-likelihood function of the BRM given in \eqref{lin:pred} can be written as
\begin{equation}\label{log-like}
    l(\bbeta)=\sum_{i=1}^n l_i(\mu_i,\phi)
\end{equation}
where
\begin{eqnarray}\label{logg}
    l_i(\mu_i,\phi) &=& \log(\Gamma(\phi))-\log(\Gamma(\mu_i\phi))-\log(\Gamma((1-\mu_i)\phi))+(\phi\mu_i-1)\log(y_i)\\&&+((1-\mu_i)\phi-1)\log(1-y_i)\nonumber.
\end{eqnarray}
Since the log-likelihood function is nonlinear in $\bbeta$, one needs to make use of an iterative procedure to obtain the MLEs of parameters. 

Therefore, we need the score function by differentiating the log-likelihood function with respect to the unknown parameters $\bbeta$ and $\phi$ respectively as
\begin{eqnarray}\label{dif:beta}
    U_{\bbeta}(\bbeta, \phi) = \phi \bX^\prime \bT(\by^* - \bmu^*)
\end{eqnarray}
and
\begin{eqnarray}\label{dif:phi}
    U_{\phi}(\bbeta, \phi) = \sum_{i=1}^n \{ \mu_i(y_i^*-\mu_i^*)+\log(1-y_i)-\psi(\phi(1-\mu_i)
    +\psi(\phi))\}
\end{eqnarray}
where $y_i^*=\log(y_i/(1-y_i)$, $\mu_i^*=\psi(\mu_i \phi)$, $\bT=diag\{1/g'(\mu_1),\ldots, 1/g'(\mu_n)  \}$, $\by^*=(y_1^*,\ldots, y_n^*)^\prime$, $\bmu^*=(\mu_1^*,\ldots, \mu_n^*)^\prime$
and the Fisher's information matrix as

It is known that under usual regularity conditions, the maximum likelihood estimators $\mle$, $\widehat{\phi}$ of $\bbeta$ and $\phi$ are asymptotically normal as $n \to \infty$,  
\begin{eqnarray}\label{dist:mle}
    \begin{pmatrix}
\mle\\
\widehat{\phi}
\end{pmatrix} \sim  N_{p+1}
\left[ \begin{pmatrix}
\bbeta\\
\phi
\end{pmatrix}, \bK^{-1} \right],
\end{eqnarray}
where 
\begin{eqnarray}\label{fisher}
    \bK=\bK(\bbeta, \phi)= \begin{pmatrix}
\bK_{\bbeta\bbeta} & \bK_{\bbeta\phi}\\
\bK_{\phi\bbeta} & \bK_{\phi\phi},
\end{pmatrix}
\end{eqnarray}
$\bK_{\bbeta\bbeta}=\phi \bX^\prime\bW\bX = \boldsymbol{\mathcal{I}}$, $\bK_{\bbeta\phi}=\bK_{\phi\bbeta}=\bX^\prime \bT$ and $\bK_{\phi\phi}=trace(\bD)$, $\bD=diag\{d_1, \ldots, d_n \}$ with $d_i=\psi^\prime(\mu_i\phi)\mu_i^2+\psi^\prime((1-\mu_i)\phi)(1-\mu_i)^2-\psi^\prime(\phi)$ such that $\psi^\prime(.)$ is the trigamma function, $\bc=(c_1, \ldots, c_n)$ with $c_i=\phi\{\psi^\prime(\mu_i\phi)\mu_i-\psi^\prime((1-\mu_i)\phi)(1-\mu_i) \}$ ( \citet{Ferrari}).

One estimation method for handling the multicollinearity problem in BRM is the ridge estimator introduced by \citet{Qasim} as follows:
\begin{equation}\label{equ3}
\widehat{\boldsymbol{\beta}}^{UR} = ( \textbf{X}^\prime  \widehat{\bW} \textbf{X} + k \textbf{I}_p )^{-1} \textbf{X}^\prime  \widehat{\bW} \bX \mle ,
\end{equation}
where $ k > 0 $ is the ridge parameter, and $\Wh $ is a diagonal matrix such that the ith diagonal element is equal to $ \widehat{\mu}_i  = exp( \bx^\prime_i \mle) $. We will call this as the unrestricted ridge estimator of the BRM, and hence the superscript "UR" comes into the picture.  The ridge parameter is a tuning parameter and often unknown. In this work, we will adopt the proposal of \citet{Qasim}  and  estimate the ridge parameter $k$ by $\widehat{k} =\frac{ 1 }{ \hat{\boldsymbol{\varphi}}^\prime \hat{\boldsymbol{\varphi}} } $ where $ \hat{\boldsymbol{\varphi}} = \textbf{C}^\prime \widehat{ \boldsymbol{\beta} } $ such that the columns of $\bC$ are the eigenvectors of the matrix $ \bX' \Wh  \bX$.

In contrast to the unrestricted estimator above, when there is some prior information regarding the parameters such as a linear restriction of the form  
\begin{equation}\label{hypo}
\textbf{H}_0 : \textbf{H} \boldsymbol{\beta} = \textbf{h} \,\, vs \,\, \textbf{H}_1 : \textbf{H} \boldsymbol{\beta} \neq \textbf{h}
\end{equation}
where $ \textbf{H} $ is a $ p_2 \times p $ matrix, $ p_2 $ is the number of non-significant parameters, and $ \textbf{h} $ is a $ p_2 \times 1 $ known vector, one can produce a restricted MLE as  
\begin{equation}\label{rmle}
\rmle = \mle  - \boldsymbol{\mathcal{I}}^{-1} \bH^\prime \Big( \bH\, \boldsymbol{\mathcal{I}} ^{-1} \textbf{H}^\prime \Big)^{-1} \Big(  \textbf{H} \, \mle - \textbf{h} \Big), 
\end{equation}
where $\boldsymbol{\mathcal{I}}^{-1}$ is the inverse of the Fisher's information matrix given previously. 

As a consequence, restricted ridge estimator can be defined as  
\begin{equation}\label{equ10}
\rr = ( \bX^\prime  \Wh \bX + k\, \bI_p )^{-1} \bX^\prime \Wh \bX \rmle,
\end{equation} \citep{kibria2012}.

Although the restriction in (\ref{hypo}) is quite general, an important special case occurs when the matrix $ \textbf{H} $  is a contrast matrix designed to display the fact that a portion of the regression effects are weak or insignificant. Such a situation is equivalent to partitioning the $\bbeta$ into $(\bbeta_1, \bbeta_2)$ and setting $H_0:\bbeta_2=0$. For instance, when a model selection has been performed and a subset of regression parameters, $\bbeta_2$ is claimed to be negligible, then one obtains the above restricted maximum likelihood  and ridge estimators for the reduced model with only the remaining set of coefficients, $\bbeta_1$.  

\subsection{The proposed estimators}
The main idea behind the estimators that we are about to propose in this section is that the restricted ridge estimators of the previous section can be improved in terms of efficiency by using  shrinkage estimation techniques (\citet{Mandal2019},  \citet{Lisawadi2020}, \citet{hossain2018}, and  \citet{hossain2012}). We will call this class of estimators as ridge-type  shrinkage estimators. The essence of these estimators is to combine the unrestricted and restricted estimators of the previous section as follows:
 \begin{itemize}
     \item {\bf Ridge--type Linear Shrinkage Estimator:}  The ridge--type linear shrinkage estimator of $ \boldsymbol{\beta} $ denoted by $ \rls$ is defined as
\begin{equation}\label{rls}
\rls = \delta \, \rr + (1 - \delta) \, \ur,
\end{equation}
where $ 0 \leq \delta \leq 1 $ is the confidence level in prior information and its optimum value is obtained by minimizing the $MSE$ of the estimator.

\item {\bf Ridge--type Pretest Estimator:}
The ridge--type pretest estimator of $ \boldsymbol{\beta} $ denoted by $ \rpt$ has the following form
\begin{equation}\label{equ13}
\rpt = \ur  - (\ur  - \rr) \, I( T_n \leq  T_{n,\alpha}) ,
\end{equation}
where $ I(.) $ is an indicator function and $ T_{n,\alpha} $ is the $ \alpha $-level upper value of the distribution of a  test statistic $ T_n $ for testing the prior information contained in the hypotheses (\ref{hypo}). The ridge--type pretest estimator has two choices so that, if $ \textbf{H}_0 : \textbf{H} \boldsymbol{\beta} = \textbf{h} $ is true then, $ \rpt = \rr $ otherwise, $ \rpt = \ur$.

In this work, we will employ the following Wald-type test statistic:
\begin{align}\label{equ11}
T_n =    n \, (\textbf{H} \ur - \textbf{h})^\prime \Big ( \textbf{H} \  ( \frac{1}{n} \boldsymbol{\mathcal{I}} )^{-1}  \textbf{H}^\prime \Big )^{-1}  (\textbf{H} \ur  - \textbf{h}).
\end{align}
As $ n \to \infty $, the above test statistic has asymptotic chi-square distribution with $ p_2 $ degrees of freedom.

\item {\bf Ridge--type Shrinkage Pretest Estimator:}
The ridge--type shrinkage pretest estimator of $ \boldsymbol{\beta} $ denoted by $ \spe$ is as
\begin{equation}\label{equ14}
\spe = \ur  - \delta \,(\ur  - \rr  ) \, I( T_n \leq  T_{n,\alpha}) ,
\end{equation}
note that, $ \spe $ is more efficient than $ \widehat{\boldsymbol{\beta}}^{RPT} $ in many parts of the parameter space.

\item {\bf Ridge--type Stein Estimator:}
We denote the ridge--type Stein estimator of $ \boldsymbol{\beta} $ by $ \rs$ that combines the ridge--type unrestricted and ridge--type restricted estimator in an optimal way, dominating the ridge--type unrestricted estimator is defined as follows  
\begin{equation}\label{rs}
\rs = \rr + [ 1 - (p_2 - 2)\, T^{-1}_n ]  \Big (\ur -  \rr \Big ) ,\qquad p_2\geq 3.
\end{equation}

\item {\bf Ridge--type Positive Stein Estimator:}
The ridge--type positive Stein estimator of $ \boldsymbol{\beta} $ denoted by $ \rps$ is defined as 
\begin{equation}\label{rps}
\rps =\rr + [ 1 - (p_2 - 2)\, T^{-1}_n ]^{+}  \Big (\ur -  \rr \Big ) ,\qquad p_2\geq 3 ,
\end{equation}
where $ z^{+} = \max (0 , z) $. The $ \rps $ adjust controls for the over--shrinking problem in ridge--type Stein estimator.
\end{itemize}


\section{Asymptotic Properties}\label{sec3}
To explore the asymptotic properties of the ridge--type shrinkage estimators introduced in Section \ref{sec2}, it is common to consider the following sequence of local alternatives, 
\begin{equation}\label{equ17}
\mathcal{K}_{(n)} \, : \,  \textbf{H}\boldsymbol{\beta} = \textbf{h} + \frac{\boldsymbol{\vartheta}}{\sqrt{n}} ,
\end{equation}
where $ \boldsymbol{\vartheta} = (\vartheta_1, \vartheta_2, ..., \vartheta_{p_2})^\prime \in R^{p_2} $ is a $ p_2 \times 1 $ vector of fixed values. In order to compare the estimators, we compute the asymptotic distributional bias $ ( \mathcal{B}) $ and variances $ (\mathcal{V}) $ of the proposed estimators \citep{Saleh2006}.

To this end, let  $ \widehat{\bbeta} $ denote any one of the estimators defined in the previous section. The asymptotic distributional bias and variance are defined, respectively, as:  
\begin{equation}\label{equ18}
\mathcal{B} ( \widehat{\bbeta}) = \lim_{n \to \infty} E \Big ( \sqrt{n} ( \widehat{\bbeta} - \boldsymbol{\beta} ) \Big ),
\end{equation}
 and  
\begin{equation}\label{equ19}
\mathcal{V} ( \widehat{\bbeta}) = \lim_{n \to \infty} E \Big ( \sqrt{n} ( \widehat{\bbeta} - \boldsymbol{\beta} )   \,   \sqrt{n} ( \widehat{\bbeta} - \boldsymbol{\beta} ) ^\prime \Big ).
\end{equation}

Although quite complicated, analytical expressions can be computed for this bias and variance quantities. Such expressions are contained in the following two theorems.  
\begin{Theorem}\label{Theorem 1} 
Under the sequence of local alternatives given in \eqref{equ17} and the usual regularity conditions, the asymptotic distributional biases of the proposed estimators are as follows,
\begin{align}
\mathcal{B} ( \ur )  & = \, ( \textbf{A} - \mathbf{I}_p )\,\boldsymbol{\beta} , \nonumber  \\
\mathcal{B} ( \rr) & =  ( \textbf{I}_p - \boldsymbol{\mathcal{J}} \,  \mathbf{H} ) ( \textbf{A} - \textbf{I}_p ) \,\boldsymbol{\beta}  - \boldsymbol{\mathcal{J}} \,   \boldsymbol{\vartheta}  , 
 \nonumber  \\
\mathcal{B}  ( \rls )  & =  \mathcal{B} ( \ur )  - \delta \,\boldsymbol{\mathcal{J}}\, [ \textbf{H}   ( \textbf{A} - \textbf{I}_p ) \,\boldsymbol{\beta} + \boldsymbol{\vartheta} ] ,
 \nonumber  \\
\mathcal{B}  ( \rpt)  & =  \mathcal{B} ( \ur ) -  \boldsymbol{\mathcal{J}}\, [ \textbf{H}   ( \textbf{A} - \textbf{I}_p ) \,\boldsymbol{\beta} + \boldsymbol{\vartheta} ]\,\boldsymbol{ \Psi}_{p_2 + 2} ( \chi^2_{p_2, \alpha} ; \Delta^{*})  ,
 \nonumber  \\
\  \  \   \mathcal{B}  ( \spe ) & =  \mathcal{B} ( \ur ) -  \delta \,\boldsymbol{\mathcal{J}}\, [ \textbf{H}   ( \textbf{A} - \textbf{I}_p ) \,\boldsymbol{\beta} + \boldsymbol{\vartheta} ] \,\boldsymbol{ \Psi}_{p_2 + 2} ( \chi^2_{p_2, \alpha} ; \Delta^{*}) ,
\nonumber  \\
\mathcal{B}  ( \rs ) & =  \mathcal{B} ( \ur ) - ( p_2 - 2 ) \, \boldsymbol{\mathcal{J}} \ [ \textbf{H} \,( \textbf{A} - \textbf{I}_p )\, \boldsymbol{\beta}   + \boldsymbol{\vartheta} ] \, E  \Big[  \frac{1}{\chi^2_{ p_2 + 2 } ( \Delta^{*} )}  \Big] ,
 \nonumber  \\
\  \   \mathcal{B}  ( \rps )  & = \mathcal{B} ( \rs ) - \, \boldsymbol{\mathcal{J}}\, [ \textbf{H} (  \textbf{A} - \textbf{I}_p ) \,\boldsymbol{\beta} + \boldsymbol{\vartheta} ] 
\bigg \lbrace  \boldsymbol{\Psi}_{p_2 + 2}(\chi^2_{p_2 ,\alpha} ; \Delta^{*} ) +
\nonumber \\
& \qquad \qquad \qquad \qquad
( p_2 - 2 ) \, E \biggl [ \dfrac{ I ( \chi^2_{p_2 + 2} (\Delta^{*}) < p_2 - 2 ) }{ \chi^2_{p_2 + 2} ( \Delta^{*} ) } \biggl ] \bigg \rbrace. 
\nonumber 
\end{align}
where $ \boldsymbol{\Psi}_{v}(. ; \Delta^{*} ) $ is the cumulative distribution function of the $ \chi^2_{v}( \Delta^{*} ) $ distribution and $ \Delta^{*} = \boldsymbol{\vartheta}^\prime ( \textbf{H} \, \textbf{I}^{-1} \textbf{H}^\prime )^{-1}  \boldsymbol{\vartheta} $ is the non-centrality parameter.
\end{Theorem}
\textbf{Proof}: See Appendix 2.

\begin{Theorem} \label{Theorem 2}
Under the local alternatives given in \eqref{equ17} and the usual regularity conditions, the asymptotic distributional variances of the estimators are as follows
\begin{align*}
\mathcal{V} ( \ur ) & =   \textbf{A} \boldsymbol{\mathcal{I}}^{-1} \textbf{A}^\prime + \bigg[ ( \textbf{A} - \mathbf{I}_p ) \,\boldsymbol{\beta} \bigg] \, \bigg[ ( \textbf{A} - \mathbf{I}_p ) \, \boldsymbol{\beta} \bigg]^\prime ,
 \nonumber  \\
\mathcal{V} ( \rr ) & =   \textbf{A} \, \boldsymbol{\mathcal{I}}^{-1} \, \textbf{A}^\prime - \boldsymbol{\mathcal{J}} \,  \mathbf{H} \, \textbf{A} \, \boldsymbol{\mathcal{I}}^{-1} \, \textbf{A}^\prime  
\nonumber \\
& \qquad + \bigg[  ( \mathbf{I}_p  - \boldsymbol{\mathcal{J}}  \,  \mathbf{H} ) \, ( \textbf{A} - \textbf{I}_p ) \, \boldsymbol{\beta} - \boldsymbol{\mathcal{J}} \,\boldsymbol{\vartheta}  \bigg] \, \bigg [  ( \mathbf{I}_p  - \boldsymbol{\mathcal{J}}  \,  \mathbf{H} ) \, ( \textbf{A} - \textbf{I}_p ) \, \boldsymbol{\beta} - \boldsymbol{\mathcal{J}} \,\boldsymbol{\vartheta} \bigg ]^\prime ,
 \nonumber  
\end{align*}

\begin{align*}
\mathcal{V} ( \rls )  & = \mathcal{V} ( \ur )  
\nonumber \\
& - 2 \delta \, \big \lbrace ( \textbf{I}_p - \boldsymbol{\mathcal{J}}\, \textbf{H} ) \, ( \textbf{A} - \textbf{I}_p ) \, \boldsymbol{\beta} ) - \boldsymbol{\mathcal{J}} \, \boldsymbol{\vartheta} \big \rbrace
\big \lbrace  \boldsymbol{\mathcal{J}}\, \textbf{H}\,( \textbf{A} - \textbf{I}_p ) \, \boldsymbol{\beta} + \boldsymbol{\mathcal{J}} \, \boldsymbol{\vartheta} \big \rbrace ^\prime 
\nonumber \\
& - \delta \, ( 2 - \delta ) \, \big \lbrace   \boldsymbol{\mathcal{J}}\, \textbf{H}\, \textbf{A}  \,  \boldsymbol{\mathcal{I}}^{-1} \, \textbf{A}^\prime +  [ \boldsymbol{\mathcal{J}}\, \textbf{H}\,( \textbf{A} - \textbf{I}_p ) \, \boldsymbol{\beta} + \boldsymbol{\mathcal{J}} \, \boldsymbol{\vartheta}  ]
\nonumber \\
 & \times [ \boldsymbol{\mathcal{J}}\, \textbf{H}\,( \textbf{A} - \textbf{I}_p ) \, \boldsymbol{\beta} + \boldsymbol{\mathcal{J}} \, \boldsymbol{\vartheta}  ]^\prime \big \rbrace  ,
 \nonumber  
\end{align*}

\begin{align*}
\mathcal{V} ( \rpt )  & = \mathcal{V} ( \ur )
- 2 \, \bigg ( 
\bigg \lbrace ( \textbf{I}_p - \boldsymbol{\mathcal{J}} \, \textbf{H}  )\, ( \textbf{A} - \textbf{I}_p ) \, \boldsymbol{\beta}  - \boldsymbol{\mathcal{J}} \, \boldsymbol{\vartheta} \bigg \rbrace 
\nonumber \\
& \times  [ \boldsymbol{\mathcal{J}} \, \textbf{H} \, ( \textbf{A} - \textbf{I}_p ) \, \boldsymbol{\beta}  + \boldsymbol{\mathcal{J}} \, \boldsymbol{\vartheta}  ]^\prime \, \Psi_{p_2 + 2} ( \chi^2_{p_2, \alpha} ; \Delta^{*} )  \bigg ) 
\nonumber \\
& -  \bigg (  
\boldsymbol{\mathcal{J}} \, \textbf{H} \, \textbf{A} \, \boldsymbol{\mathcal{I}}^{-1} \, \textbf{A}^\prime \, \Psi_{p_2 + 2} ( \chi^2_{p_2, \alpha} ; \Delta^{*} )  + [ \boldsymbol{\mathcal{J}} \, \textbf{H} \, ( \textbf{A} - \textbf{I}_p ) \, \boldsymbol{\beta}  + \boldsymbol{\mathcal{J}} \, \boldsymbol{\vartheta} ] 
\nonumber \\
 & \times [ \boldsymbol{\mathcal{J}} \, \textbf{H} \, ( \textbf{A} - \textbf{I}_p ) \, \boldsymbol{\beta}  + \boldsymbol{\mathcal{J}} \, \boldsymbol{\vartheta} ]^\prime  \Psi_{p_2 + 4} ( \chi^2_{p_2, \alpha} ; \Delta^{*} )
\bigg) , 
 \nonumber  
\end{align*}

\begin{align*}
\mathcal{V} ( \spe ) & =  \mathcal{V} ( \ur )
 - 2 \delta \, \bigg ( 
\lbrace ( \textbf{I}_p - \boldsymbol{\mathcal{J}} \, \textbf{H}  )\, ( \textbf{A} - \textbf{I}_p ) \, \boldsymbol{\beta}  - \boldsymbol{\mathcal{J}} \, \boldsymbol{\vartheta} \rbrace 
\nonumber \\
& \times [ \boldsymbol{\mathcal{J}} \, \textbf{H} \, ( \textbf{A} - \textbf{I}_p ) \, \boldsymbol{\beta}  + \boldsymbol{\mathcal{J}} \, \boldsymbol{\vartheta}  ]^\prime \, \Psi_{p_2 + 2} ( \chi^2_{p_2, \alpha} ; \Delta^{*} )  \bigg ) 
\nonumber \\
& - \delta \, ( 2 - \delta ) \, 
\bigg (  
\boldsymbol{\mathcal{J}} \, \textbf{H} \, \textbf{A} \, \boldsymbol{\mathcal{I}}^{-1} \, \textbf{A}^\prime \, \Psi_{p_2 + 2} ( \chi^2_{p_2, \alpha} ; \Delta^{*} )  + [ \boldsymbol{\mathcal{J}} \, \textbf{H} \, ( \textbf{A} - \textbf{I}_p ) \, \boldsymbol{\beta}  + \boldsymbol{\mathcal{J}} \, \boldsymbol{\vartheta} ] 
\nonumber \\
& \times [ \boldsymbol{\mathcal{J}} \, \textbf{H} \, ( \textbf{A} - \textbf{I}_p ) \, \boldsymbol{\beta}  + \boldsymbol{\mathcal{J}} \, \boldsymbol{\vartheta} ]^\prime  \Psi_{p_2 + 4} ( \chi^2_{p_2, \alpha} ; \Delta^{*} )
\bigg) , 
 \nonumber  
\end{align*}
 
\begin{align*}
\mathcal{V} ( \rs )   & = \mathcal{V} ( \ur )
 \nonumber \\
& - 2 \, ( p_2 - 2 ) \, \bigg (
\big [ ( \textbf{I}_p - \boldsymbol{\mathcal{J}} \, \textbf{H} ) \, ( \textbf{A} - \textbf{I}_p ) \,\boldsymbol{\beta} - \boldsymbol{\mathcal{J}} \, \boldsymbol{\vartheta} \big]\,
 \nonumber \\
& \times \big [ \boldsymbol{\mathcal{J}} \, \textbf{H} [ \textbf{A} - \textbf{I}_p ] \, \boldsymbol{\beta} + \boldsymbol{\mathcal{J}} \, \boldsymbol{\vartheta} \big ] \, E  \Big [ \frac{1}{\chi^2_{p_2 + 2}(\Delta^{*})} \Big ]
\bigg )
 \nonumber \\
& + ( p_2 - 2 ) \, ( p_2 - 4 )\, 
\boldsymbol{\mathcal{J}} \, \textbf{H} \, \textbf{A} \,  \boldsymbol{\mathcal{I}}^{-1} \, \textbf{A}^\prime \, 
\bigg (  
E  \Big [ \frac{1}{(\chi^2_{p_2 + 2}(\Delta^{*}))^2} \Big ] - E  \Big [ \frac{1}{\chi^2_{p_2 + 2}(\Delta^{*})} \Big ]
\bigg )
\nonumber \\
& + ( p_2 - 2 ) \, ( p_2 - 4 )\, 
[ \boldsymbol{\mathcal{J}} \, \textbf{H}  \, ( \textbf{A} - \textbf{I}_p ) \,\boldsymbol{\beta} + \boldsymbol{\mathcal{J}} \, \boldsymbol{\vartheta} ] \,
[ \boldsymbol{\mathcal{J}} \, \textbf{H}  \, ( \textbf{A} - \textbf{I}_p ) \,\boldsymbol{\beta} + \boldsymbol{\mathcal{J}} \, \boldsymbol{\vartheta} ]^\prime 
\nonumber \\
& \times \bigg (  
E  \Big [ \frac{1}{(\chi^2_{p_2 + 4}(\Delta^{*}))^2} \Big ] - E  \Big [ \frac{1}{\chi^2_{p_2 + 4}(\Delta^{*})} \Big ]
\bigg ) ,
 \nonumber  
\end{align*}
 
\begin{align*}
\mathcal{V} ( \rps )  & =  \mathcal{V} ( \rs ) 
\nonumber \\
& - 2\, \bigg (
 \lbrace ( \textbf{I}_p -  \boldsymbol{\mathcal{J}} \, \textbf{H} ) \, ( \textbf{A} - \textbf{I}_p ) \,\boldsymbol{\beta}  - \boldsymbol{\mathcal{J}} \, \boldsymbol{\vartheta} \rbrace \,
 [ \boldsymbol{\mathcal{J}} \, \textbf{H}  \, ( \textbf{A} - \textbf{I}_p ) \,\boldsymbol{\beta}  - \boldsymbol{\mathcal{J}} \, \boldsymbol{\vartheta} + \boldsymbol{\mathcal{J}} \, \boldsymbol{\vartheta} ] \, 
\nonumber \\
& \times E  \Big [ \Big (1 - \frac{p_2 - 2}{\chi^2_{p_2 + 2}(\Delta^{*})} \Big ) \, I ( \chi^2_{p_2 + 2}(\Delta^{*}) <  p_2 - 2 )\Big ]
\bigg )
 \nonumber \\
& - \bigg (
\boldsymbol{\mathcal{J}} \, \textbf{H}  \,  \textbf{A} \,  \boldsymbol{\mathcal{I}}^{-1}\, \textbf{A}^\prime
\, E\Big [ \Big( 1 - \frac{p_2 -2}{\chi^2_{p_2 + 2}(\Delta^{*})} \Big )^2 \, I ( \chi^2_{p_2 + 2}(\Delta^{*}) <  p_2 - 2 )  \Big ]
\nonumber \\
&  + [ \boldsymbol{\mathcal{J}} \, \textbf{H}  \, ( \textbf{A} - \textbf{I}_p ) \,\boldsymbol{\beta} + \boldsymbol{\mathcal{J}} \, \boldsymbol{\vartheta} ] \, 
[ \boldsymbol{\mathcal{J}} \, \textbf{H}  \, ( \textbf{A} - \textbf{I}_p ) \,\boldsymbol{\beta}  + \boldsymbol{\mathcal{J}} \, \boldsymbol{\vartheta} ]^\prime
\nonumber \\
&  \times E\Big [ \Big( 1 - \frac{p_2 -2}{\chi^2_{p_2 + 4}(\Delta^{*})} \Big )^2 \, I ( \chi^2_{p_2 + 4}(\Delta^{*}) <  p_2 - 2 )  \Big ]
\bigg ) .
\nonumber
\end{align*}
\end{Theorem}
\textbf{Proof}: See Appendix 3.


\section{Monte Carlo Simulation}\label{sec4} 
Based on the results obtained in the previous section, it's clear that the theoretical properties of the proposed estimators are not in closed forms and the computation of the biases and variances depend on several unknown parameters. Therefore, we employ a Monte Carlo simulation to study the performance of the proposed estimators in terms of their  relative mean square errors (RMSE). 
\subsection{Low Dimensional Data }
To this end, we generate a set of  predictor variables  from a $p-dimensional$ multivariate normal distribution with zero means $\bf 0$ and a covariance matrix $\boldsymbol{\Sigma}$, where $\Sigma_{ij}=\rho^{|i-j|}$, $i,j=1,2, \ldots, p$, and  $\rho$ controls the degree of correlation between the predictors, and it is taken as $0.6$ and $0.9$.

\begin{figure}
    \centering
    \includegraphics[width=\textwidth]{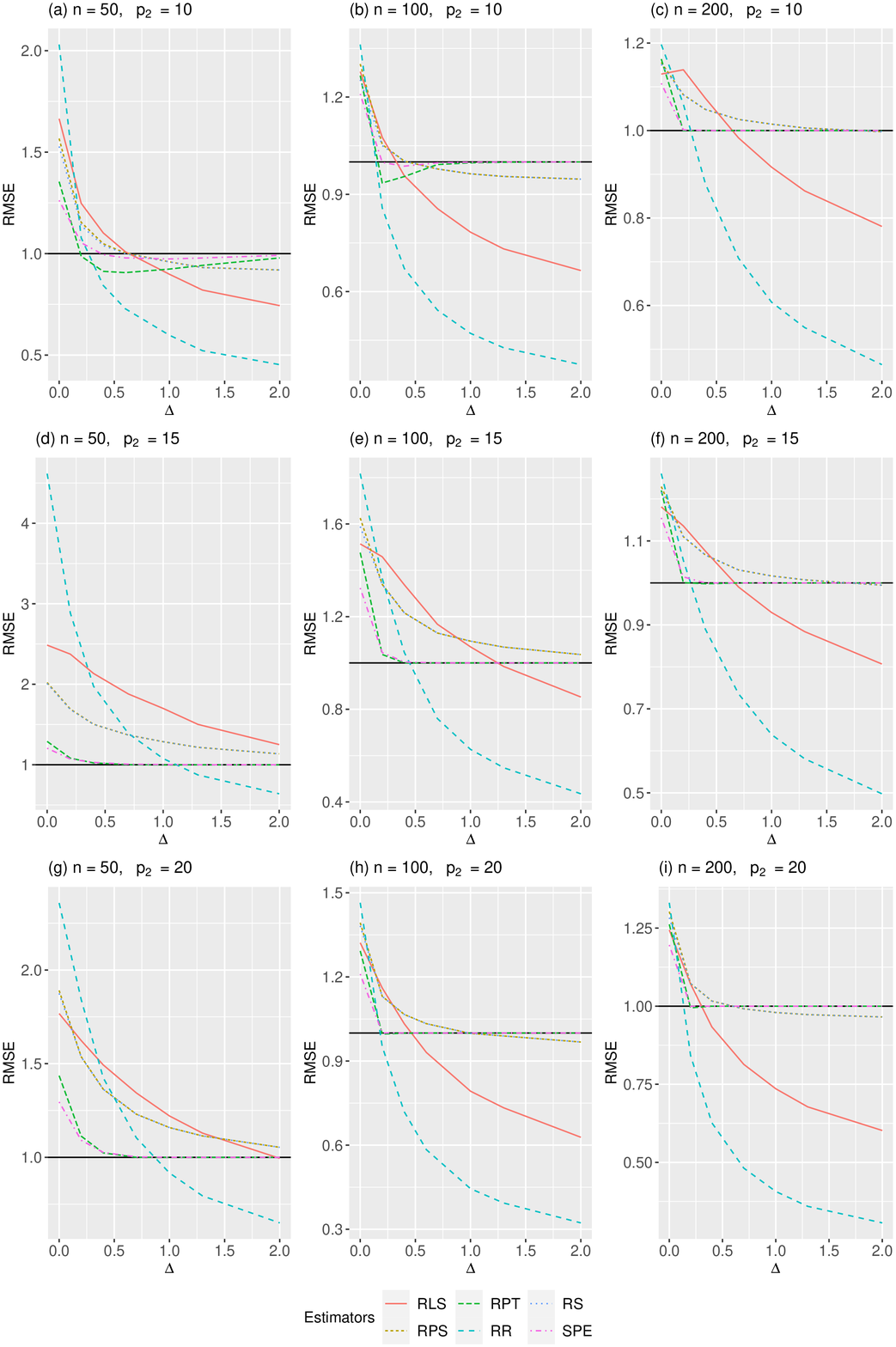}
    \caption{RMSEs of the estimators versus $\Delta$ when $\rho=0.6$ in low dimensional setting with different values of $p_2$ and $n$.}
    \label{rho06}
\end{figure}

\begin{figure}
    \centering
    \includegraphics[width=\textwidth]{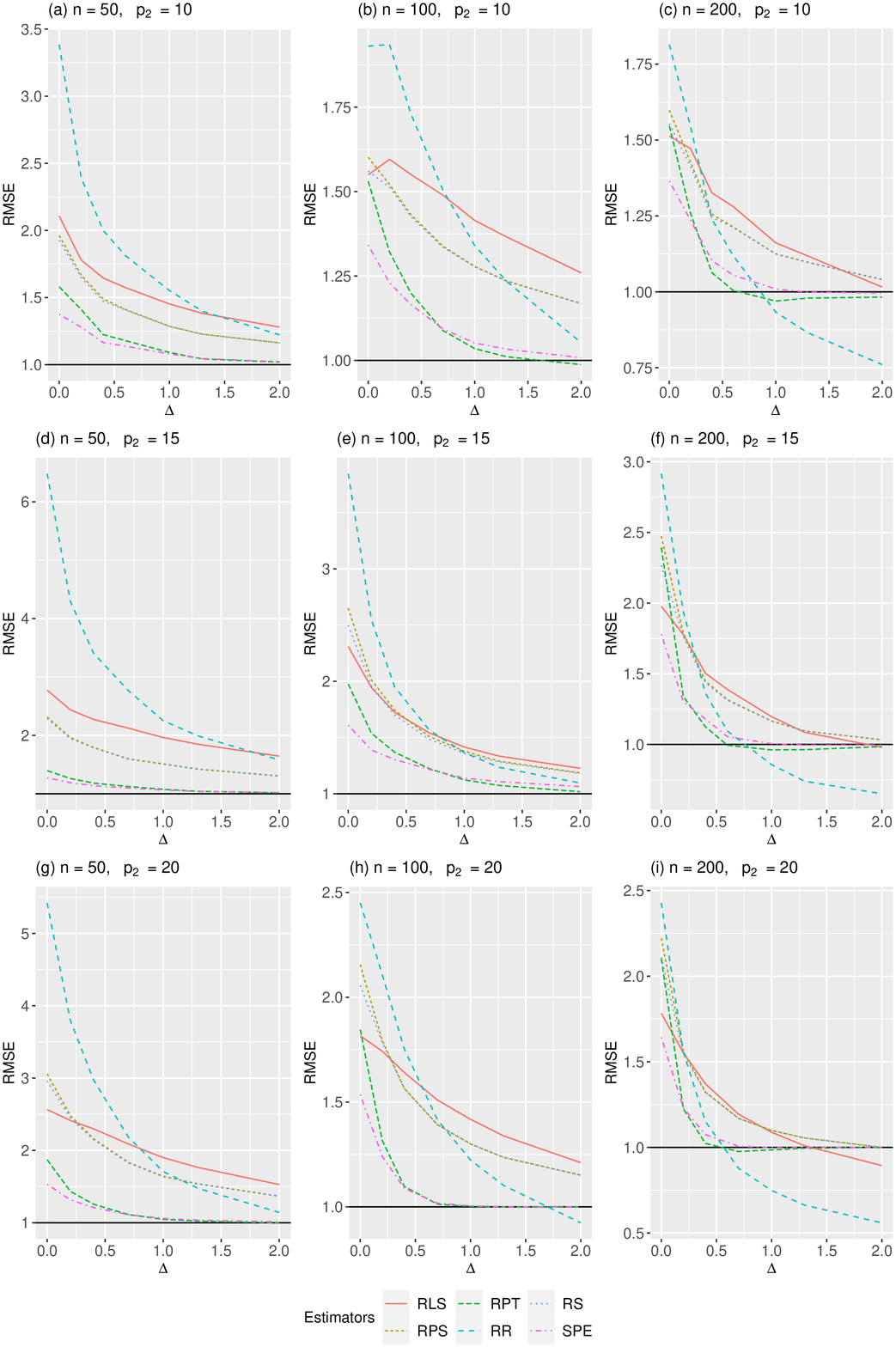}
    \caption{RMSEs of the estimators versus $\Delta$ when $\rho=0.9$ in low dimensional setting with different values of $p_2$ and $n$.}
    \label{rho09}
\end{figure}
\begin{figure}
    \centering
    \includegraphics[width=\textwidth]{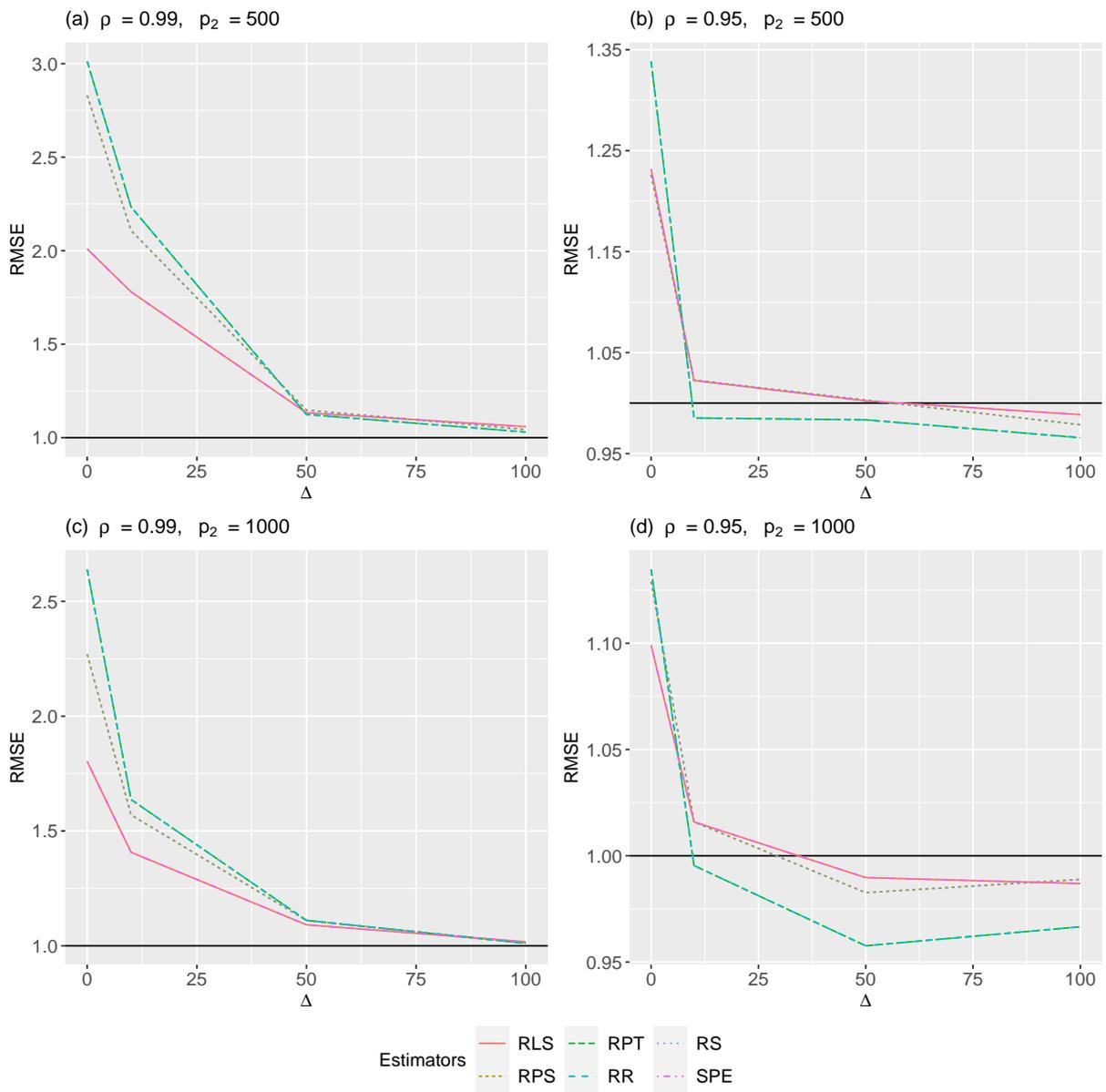}
    \caption{RMSEs of estimators versus $\Delta$ in high dimensional setting with different values of $p_2$ and $\rho$.}
    \label{fig:HD}
\end{figure}

The matrix  $ \textbf{H} $ that defined the restriction in (\ref{hypo}) is chosen as  $\textbf{H}=\left(\bzero_{p_2 \times p_1}, \bI_{p_2} \right) \in \mathbb{R}^{p_2 \times p}$,  a matrix of rank $p_2$, and $\bI_{p_2}\in \mathbb{R}^{p_2 \times p_2}$ is an identity matrix of order $p_2$ such that $p=p_1+p_2$. As we have explained earlier, this is equivalent to assuming that the regression parameter vector is partitioned as $\bbeta=\left(\bbeta_1^\prime, \bbeta_2^\prime \right)^\prime$ where $\bbeta_1 \in \mathbb{R}^{p_1}$ and $\bbeta_2 \in \mathbb{R}^{p_2}$ are, respectively, the active and the inactive parameter vectors.

The outcome is assumed to follow a beta distribution and therefore, its mean, $\mu_i$ is linked to the regression parameters via a logit link,   
\begin{equation}
    \mu_i=\frac{\exp(\bx_i^\prime\bbeta)}{1+\exp(\bx_i^\prime\bbeta)}. \nonumber
\end{equation}

Using these $\mu_i$ and by setting the dispersion parameter of the  beta distribution as $\phi=5$, we  generate the outcome as   $y_i \sim Beta\left( \mu_i \phi, (1-\mu_i)\phi\right)$. 
We considered sample sizes of  $n=50, 100, 200$ while the 
 number of inactive variables, $(p_2)$, is  varied over the set  $p_2=10, 15, 20$. 

The next parameter to be varied is the quality of the prior information. That is the distance between the true  $\bbeta_2$ and the vector $\bzero_{p_2}$, $\Delta=\lVert \bbeta-\bbeta^{(0)} \rVert$ where $\lVert . \rVert$ is the usual Euclidean norm and $\bbeta^{(0)}=\left(\bbeta_1^\prime, \bzero_{p_2}^\prime \right)^\prime$ is the parameter vector under $\textbf{H}_0$. When   $\Delta=0$, the prior information (or in other words, the reduced model) is the true model while, the   more we move $\Delta$ away from zero the more  the full model with $\bbeta_2 \neq \bzero_{p_2}$ becomes the truth and the restricted estimators become more and more inefficient. Therefore, in this simulation, we choose $\bbeta_1^\prime=(2.75, -1.75, 1.45)$, while $\bbeta_2=\left(\sqrt{\Delta}, \bzero_{p_2-1} ^\prime \right)^\prime$ with $\Delta$ varying in the interval $[0,2]$.  Each scenario of the simulation is repeated 1000 times. Using these 1000 repetitions, we compute the average  relative mean squared errors  with respect to $\ur$ from the formula   
\begin{eqnarray}
    RMSE(\hbbeta^*)=\frac{\sum_k^{1000} MSE_k(\ur)}{\sum_k^{1000}MSE(\hbbeta^*)}
\end{eqnarray}
where $\hbbeta^*$ is one of the proposed estimators in the previous section. An RMSE value larger than one indicates the degree of superiority of the estimator $\hbbeta^*$ over $\ur$.

\subsection{High Dimensional Data}
In this part, we conducted the simulation study for a high dimensional beta regression model. To our knowledge, there is only one \textcolor{blue}{\texttt{R}} package called \textcolor{blue}{\texttt{betaboost}} developed by \citep{Mayr2018} to handle the high dimensional beta regression. However, our simulation results showed that this algorithm has a main drawback. For highly correlated data, the betaboost algorithm has high biases (coefficients) and high variances for the number of the selected variables (see Appendix 4, for example). As a result, the betaboost is not an efficient algorithm for very high dimensional beta regression. Therefore, we used a new machine learning algorithm called Boruta which is implemented in the \textcolor{blue}{\texttt{R}} package  \textcolor{blue}{\texttt{Boruta}} \citep{Boruta} to remove the noise variables and decrease the dimension. However, the Boruta algorithm does not produce the regression coefficients. Further, it selected many noise variables. Therefore, we applied the new methodology developed in this paper for different combination of the sample size and noise variables to see the performance of the methodology we developed in this paper. When applying Boruta, we employed the  \textcolor{blue}{\texttt{TentativeRoughFix()}} function to select the tentative variables.

In high dimensional setting, we considered the following scenarios: $n=200$, $p_2 = 500, 1000$ while $p_1 = 10$. 
Without loss of generality, the degree of correlation  were taken as $\rho = 0.95$ and $0.99$ as in high dimension data the value of $\Sigma_{ij}$ will be very small as the dimension increases. It should also be mention here that  data generation process was similar to the low dimensional case. 

\subsection{Simulation Results}

The results of the simulation studies for low  and high dimension cases are given below:
\begin{itemize}
    \item[\textbf{a.}] \textbf{Results of low dimension case}

 Figures \ref{rho06} and \ref{rho09}. These plots depict the RMSE in their vertical axis, while the horizontal axis shows the  the parameter $\Delta$.  From these graphs, the following conclusions can be made:
\begin{itemize}
\item
The RMSE  of the estimators increase as the correlation among the  predictor variables increase.
\item
As $ p_2 $ increases, the RMSEs of the all estimators increase. Also, at $ \Delta = 0 $, the performance of the ridge--type restricted estimator is the best. When $\Delta$ moves away from zero (i.e., the reduced model is not true anymore)  the RMSE of the restricted estimator decreases sharply. 
\item
At $ \Delta = 0 $, the ridge--type positive Stein estimator is better than the ridge--type Stein estimator. However, as $ \Delta $ moves away from zero, the performance of these two estimators is the same.
\item
The ridge--type Stein and positive Stein estimators are uniformly better than the all other  estimators.
\end{itemize}

\item[\textbf{b.}] \textbf{Results of high dimension case}

The results of the high dimensional cases are given in Figure \ref{fig:HD}. 

\begin{itemize}
    \item 
    We observed that restricted ridge and ridge type shrinkage estimators have better performances in all range of $\Delta$.  
    \item
    However, unlike the low dimensional case, the proposed estimators work well in a very high range of $\Delta$. However, the pattern is similar to the low dimensional case
    \item When the correlation decreases from 0.99 to 0.95 the performance of the estimators declines.  
    
    \item 
    It should be mentioned that for the small correlations, the performance of the proposed estimators were not satisfactory, however, it was not surprising as the methodology developed here was for the highly correlated variables. 
    \item
    Further, even when $\rho$ is 0.99, still we have strong, medium and low correlations. Therefore, the high dimension setting in this paper will work for many real data application.       
\end{itemize}
\end{itemize}

\section{Real Data Application}\label{app}

In this section, we apply the proposed estimators for two real data set. 

\subsection{Low dimensional data analysis for city budget data}
This data set concerning the "Government Spending in  Dutch cities" which is contained in \textcolor{blue}{\texttt{fmlogit}} package in \textcolor{blue}{\texttt{R}} as a data frame with 429 observations and 12 variables \citep{Buis2009}.The aim is to explain the proportion of city budget that is spent   on administration and government  based on 10 covariates.

The variable \textit{governing} is the dependent variable and the rest are the explanatory variables, which are given in Table \ref{data:description}.

\begin{table}[b] 
\centering
 \caption{Variable descriptions in the data set}
 \label{data:description}
 \small
 \begin{tabular}{ll}
 \hline
 governing & proportion of budget spent on governing \\
 houseval & average value of a house in 100,000 euros.\\
 popdens  & population density in 1000s of persons per square km.\\
 noleft   & no left party in city government\\
 minorityleft & left parties are minority in city government\\
 safety & proportion of budget spent on safety \\
 education & proportion of budget spent on education \\
 recreation & proportion of budget spent on recreation \\
 social & proportion of budget spent on social work \\
 urbanplanning & proportion of  budget spent on urban planning \\
 tot & total budget in 10s of millions of euros \\
 \hline
 \end{tabular}
 \end{table}
 
\begin{figure}[h] 
\centering
\includegraphics[width=0.7\textwidth]{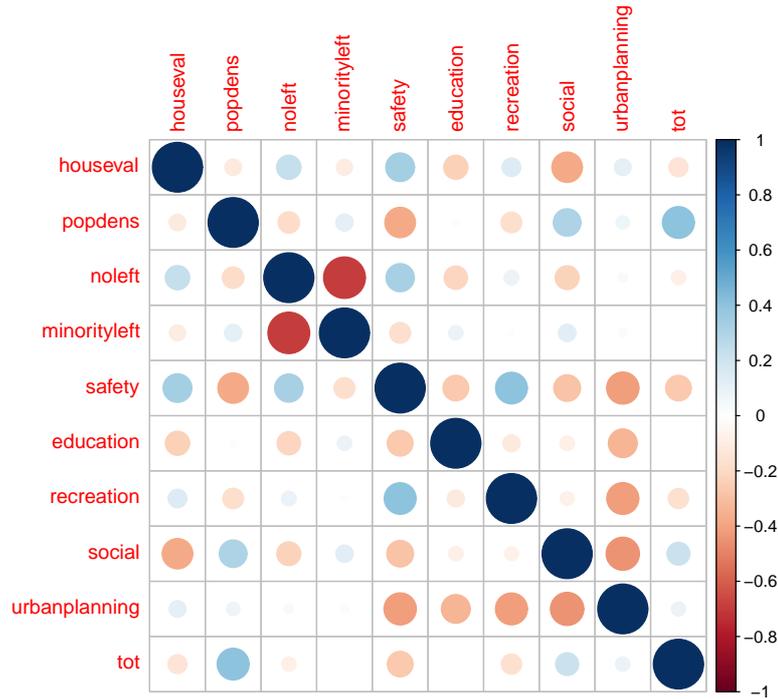}
\caption{Bivariate correlation plot of the explanatory variables in the German data.}
\label{cor.plot}
\end{figure}

\begin{table}[] 
\centering
 \caption{Coefficients and bootstrapped standard errors of the proposed estimators}
 \label{realdata}
 \small
 \begin{tabular}{lrrrrrrr}
 &\multicolumn{7}{c}{Coefficients}\\
  \hline
Variables & UR & RR & RLS & RPT & SPE & RS & RPS \\ 
  \hline
  houseval & 0.1121 & 0.1468 & 0.1294 & 0.1121 & 0.1121 & 0.1202 & 0.1202 \\ 
  education & -3.4655 & -3.5205 & -3.4930 & -3.4655 & -3.4655 & -3.4782 & -3.4782 \\ 
  recreation & -2.1481 & -1.9689 & -2.0585 & -2.1481 & -2.1481 & -2.1066 & -2.1066 \\ 
  social & -3.4400 & -3.5269 & -3.4835 & -3.4400 & -3.4400 & -3.4601 & -3.4601 \\ 
  urbanplanning & -3.6324 & -3.7032 & -3.6678 & -3.6324 & -3.6324 & -3.6488 & -3.6488 \\ 
  popdens & -0.0177 & -0.0126 & -0.0151 & -0.0177 & -0.0177 & -0.0165 & -0.0165 \\ 
  noleft & -0.0169 & -0.0227 & -0.0198 & -0.0169 & -0.0169 & -0.0182 & -0.0182 \\ 
  minorityleft & -0.1008 & -0.0488 & -0.0748 & -0.1008 & -0.1008 & -0.0888 & -0.0888 \\ 
  safety & -0.0744 & -0.3824 & -0.2284 & -0.0744 & -0.0744 & -0.1458 & -0.1458 \\ 
  tot & -0.0012 & -0.0003 & -0.0007 & -0.0012 & -0.0012 & -0.0010 & -0.0010 \\  
   \hline
   &\multicolumn{7}{c}{Standard Errors}\\
  \hline
  houseval & 0.0501 & 0.0424 & 0.0455 & 0.0522 & 0.0510 & 0.0493 & 0.0493 \\ 
  education & 0.1741 & 0.1707 & 0.1706 & 0.1775 & 0.1759 & 0.1737 & 0.1737 \\ 
  recreation & 0.2558 & 0.2161 & 0.2281 & 0.2552 & 0.2483 & 0.2444 & 0.2444 \\ 
  social & 0.1608 & 0.1251 & 0.1382 & 0.1673 & 0.1623 & 0.1566 & 0.1565 \\ 
  urbanplanning & 0.1421 & 0.1278 & 0.1322 & 0.1473 & 0.1455 & 0.1411 & 0.1411 \\ 
  popdens & 0.0219 & 0.0066 & 0.0128 & 0.0204 & 0.0195 & 0.0186 & 0.0186 \\ 
  noleft & 0.0510 & 0.0120 & 0.0263 & 0.0477 & 0.0447 & 0.0422 & 0.0422 \\ 
  minorityleft & 0.0472 & 0.0117 & 0.0247 & 0.0495 & 0.0455 & 0.0406 & 0.0406 \\ 
  safety & 0.2803 & 0.0885 & 0.1633 & 0.2869 & 0.2691 & 0.2448 & 0.2448 \\ 
  tot & 0.0042 & 0.0015 & 0.0028 & 0.0043 & 0.0042 & 0.0039 & 0.0039 \\ 
   \hline
\end{tabular}
\end{table}
Since there are missing values in some observations, we exclude them and make a complete case analysis. We fit a beta regression model and observe the significant variables. We summarize the unrestricted and restricted models in Table \ref{fm:sm}.

From the Figure \ref{cor.plot}, it is readily seen that there is a high correlation between some covariates. Also, we have computed the condition number (CN) of the matrix of cross products $\textbf{X}^\prime  \widehat{\bW} \textbf{X}$, defined as the square root of the ratio of the maximum eigenvalue to the minimum eigenvalue, such that it is computed as $809.097$. Both of the correlation plot and the CN indicate that there is a notable amount of multicollinearity among the covariates and hence the usual beta regression may not be appropriate for this data since such analysis may result in unreliable estimates. 

Therefore, we applied the proposed estimators given in this paper. By using AIC criterion, we chose a model with only five of the variables (see Table \ref{fm:sm}) as the \textit{active} variables, while the remaining five variables were deemd \textit{inactive} or week factors. The chosen variables were; \textit{houseval}, \textit{education}, \textit{recreation}, \textit{social} and \textit{urbanplanning}. Therefore, we used this restriction in the analysis and computed the proposed ridge--type shrinkage estimators and their variances (Table \ref{realdata}). We used a bootstrapping approach to compute the variances of the estimators. The results show that the bootstrap standard errors of the proposed estimators, specifically the Stein--type and positive Stein--type shrinkage estimators, are generally lower than those based on the unrestricted maximum likelihood estimator of the beta regression model. Also, the clear message from this application is that, although by using an AIC criteria to screen the variables one would incline towards a reduced model, it is worthy to recover some information from those variables deemed as weak or irrelevant. The method to recover such information is by using the shrinkage--type estimators proposed in this paper, which produce more efficient estimators in terms of reduced variances. 

\begin{table}[h] 
\centering
 \caption{Variables included in the competing models}
 \label{fm:sm}
 \small
 \begin{tabular}{lllll}
 \hline
 Method & $p_1$ & $p_2$ & Active & AIC\\
 \hline
 The full model & 10 & -- & houseval +  education + recreation + social  &-1981.154 \\
 &&& + urbanplanning + popdens + noleft \\ 
 &&& + minorityleft + safety + tot\\
 The restricted model & 5 & 5 & houseval + education + recreation  &-1978.193\\
 &&& + social + urbanplanning \\
 \hline
 \end{tabular}
 \end{table}

\subsection{\bf High dimensional data analysis for the body fat data}

The body fat data set was introduced by \citet{Weisberg1985} and it is available in \textcolor{blue}{\texttt{R}} package \textcolor{blue}{\texttt{mfp}}. The outcome variable is the percentage of body fat determined by the underwater
weighting technique. This underwater weighting technique can be inconvenient in practice since
it requires estimating the body density, which in turn requires measuring the difference of body weight measured in air and during water submersion. Hence, it is desirable to build a simple model
to estimate the percentage of body fat by using only a few measurements. The predictors in this data
set include 13 simple body measurements (such as age, weight, height and abdomen circumference). One might be interested to restrict the study for your people (Age $< 28$ years). In this case, the usual beta regression does not work as the number of predictors ($p=13$) is higher than the number of observation ($n=10$).

\begin{figure}[t] 
\centering
\includegraphics[width=0.6\textwidth, angle= -90]{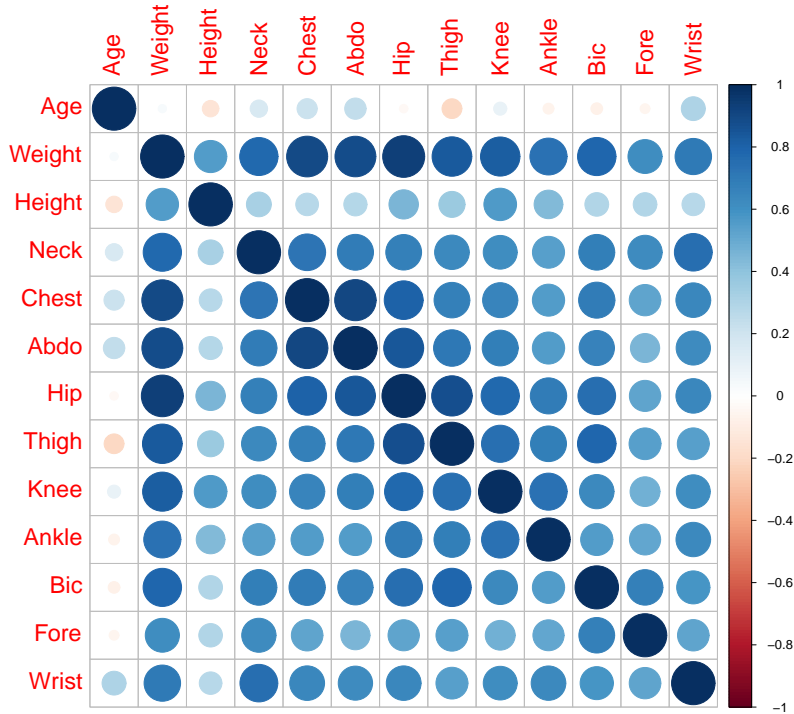}
\caption{Bivariate correlation plot of the explanatory variables in the body fat data.}
\label{cor.plot_bodyfat}
\end{figure}
 
\begin{table}[h] 
\centering
 \caption{Variables included in the competing models in body fat data}
 \label{fm1:sm1}
 \small
 \begin{tabular}{lllll}
 \hline
 Method & $p_1$ & $p_2$ & Active & AIC\\
 \hline
 The full model & 10 & -- &  Weight + Neck + Chest  &-421.4236 \\
 &&&+ Abdo + Hip + Thigh + Knee   \\ 
 &&& +           Bic + Wrist + Noise397\\
 The restricted model & 5 & 5 & Weight + Neck + Chest + Abdo + Hip   &-417.0902\\
 &&&  \\
 \hline
 \end{tabular}
 \end{table}
 
\begin{table}[] 
\centering
 \caption{Coefficients and bootstrapped standard errors of the proposed estimators body fat data}
 \label{realdata2}
 \small
 \begin{tabular}{lrrrrrrr}
 &\multicolumn{7}{c}{Coefficients}\\
  \hline
Variables & UR & RR & RLS & RPT & SPE & RS & RPS \\ 
  \hline
  Weight & 0.0259 & 0.0272 & 0.0265 & 0.026 & 0.0259 & 0.0261 & 0.0261 \\ 
Neck & -0.0416 & -0.0425 & -0.042 & -0.0417 & -0.0417 & -0.0418 & -0.0418 \\ 
Chest & -0.0141 & -0.0151 & -0.0146 & -0.0142 & -0.0142 & -0.0143 & -0.0143 \\ 
  Abdo & 0.0559 & 0.0558 & 0.0559 & 0.0559 & 0.0559 & 0.0559 & 0.0559 \\ 
  Hip & -0.0424 & -0.0432 & -0.0428 & -0.0425 & -0.0424 & -0.0425 & -0.0425 \\ 
Thigh & -0.002 & -0.0069 & -0.0045 & -0.0027 & -0.0023 & -0.0029 & -0.0029 \\ 
Knee & -0.0276 & -0.0123 & -0.0199 & -0.026 & -0.0268 & -0.0249 & -0.0249 \\ 
Bic & 0.0018 & -0.0074 & -0.0028 & 0.0011 & 0.0014 & 2e-04 & 2e-04 \\ 
Wrist & -0.0204 & -0.0147 & -0.0175 & -0.0197 & -0.02 & -0.0194 & -0.0194 \\ 
  Noise397 & 0.0218 & 2e-04 & 0.011 & 0.0196 & 0.0207 & 0.0179 & 0.0179 \\ 
   \hline
   &\multicolumn{7}{c}{Standard Errors}\\
  \hline
  Weight & 0.0054 & 0.0055 & 0.0054 & 0.0053 & 0.0053 & 0.0053 & 0.0053 \\ 
Neck & 0.0126 & 0.0121 & 0.0122 & 0.0124 & 0.0125 & 0.0124 & 0.0124 \\ 
Chest & 0.0067 & 0.0068 & 0.0067 & 0.0068 & 0.0067 & 0.0067 & 0.0067 \\ 
Abdo & 0.0069 & 0.0071 & 0.007 & 0.0069 & 0.0069 & 0.0069 & 0.0069 \\ 
Hip & 0.006 & 0.0053 & 0.0053 & 0.0058 & 0.0059 & 0.0057 & 0.0057 \\ 
Thigh & 0.007 & 0.0027 & 0.0041 & 0.0072 & 0.007 & 0.0061 & 0.0061 \\ 
Knee & 0.0079 & 0.0033 & 0.0044 & 0.0094 & 0.0085 & 0.007 & 0.007 \\ 
Bic & 0.0095 & 0.0022 & 0.005 & 0.0099 & 0.0096 & 0.0083 & 0.0083 \\ 
  Wrist & 0.0077 & 0.0028 & 0.0046 & 0.008 & 0.0078 & 0.0068 & 0.0068 \\ 
  Noise397 & 0.0108 & 0.0023 & 0.0056 & 0.0132 & 0.0118 & 0.01 & 0.01 \\ 
   \hline
\end{tabular}
\end{table}

On the other hand, there is a high correlation in between covariates which makes the analysis even more difficult. Following our discussion in the high dimensional case. Boruta selected the variables \textit{abdo} and \textit{Neck}, and therefore our methodology will not apply here. We created another scenario by adding $1000$ noise variables to our model. We applied the methodology developed here. The selected variables using Boruta are given in the first column of Table \ref{fm1:sm1}. Interestingly, Boruta selected only one noise variable and nine important variables. Further, using AIC criteria, we selected five inactive variables and applied the estimator developed in this paper. The coefficient and standard errors are given in Table \ref{fm1:sm1}. On the other hand, betaboost algorithm selected only two of the true variables (\textit{Age} and \textit{abdo}) but $36$ of the noise variables, which is not a reasonable analysis (See Table \ref{betaboostSelect} in Appendix \ref{Appendix5}). Therefore, our methods work better than the two variable selection approaches in the literature.

\section{Conclusion}\label{conc}
In this paper, we considered different types of ridge--type shrinkage estimators, namely, restricted, preliminary test, Stein--type, positive Stein--type, and linear shrinkage estimators for the low and high dimensional beta regression model. We obtained the analytical biases and variances of the proposed estimators under the local alternative hypothesis. Further, we conducted an extensive simulation study to examine the performance of the proposed estimators. Our results showed that Stein--type estimators uniformly  outperform the usual maximum likelihood estimators in the low dimensional model. Other shrinkage type estimators also had higher relative efficiencies compared with the MLEs in a wide range of parameter space. For high dimensional data, the restricted ridge and ridge shrinkage estimated performed better than the Stein type estimators.  We concluded the paper by applying the proposed methodology for two real data. The superiority of the proposed estimators were evident in terms of having less bootstrap standard errors.      
\section*{Acknowledgement}
The research is supported by Discovery Grant of the Natural Sciences and  Engineering Research Council of Canada (NSERC).


\include{Appendix}

\end{document}

%% file: Appendix.tex
\section*{Appendix 1. Auxiliary Lemma}

We present the following lemma which is  useful for the proof of the Theorems 3.1-3.3. 
\begin{Lemma} \label{lem3} 
Let $ \textbf{y} $ be a $ p_2 $-dimensional random vector distributed as $ \mathcal{N}_{p_2} ( \boldsymbol{ \mu }_{\textbf{y}} , \boldsymbol{\Sigma}_{ \textbf{y} } ) $. Then, for any measurable function $ \varphi $, we have
\begin{equation}\label{equ21}
E [ \textbf{y} \, \varphi ( \textbf{y}^\prime \textbf{y} ) ]   = \boldsymbol{ \mu }_{\textbf{y}} \, E [ \varphi ( \chi^2 _{p_2 + 2} ( \Delta^{*} ) ) ] ,
\end{equation}
\begin{equation}\label{equ22}
E [ \textbf{y}^\prime \textbf{y} \,  \varphi ( \textbf{y}^\prime \textbf{y} ) ]   = \boldsymbol{\Sigma}_{ \textbf{y} }\,  E [ \varphi ( \chi^2 _{p_2 + 2} ( \Delta^{*} ) ) ] + \boldsymbol{ \mu }_{\textbf{y}}^\prime \, \boldsymbol{ \mu }_{\textbf{y}} E [ \varphi ( \chi^2 _{p_2 + 4} ( \Delta^{*} ) ) ] ,
\end{equation}
where $ \Delta^{*} $ is the non-centrality parameter.
\end{Lemma}
\textbf{Proof:} See \citet{JudgeBock1987}.\\

\newpage
\section*{Appendix 2. Proof of Theorem 3.1}
Under local alternatives in Equation (17) we show $ \boldsymbol{\kappa}^n_1 $ is asymptotically normally distributed with mean and variance as follow
\begin{align*}
 \boldsymbol{\kappa}^n_1 & = \sqrt{n} (  \ur- \boldsymbol{\beta} )   
 \nonumber  \\
 & = \sqrt{n} ( \textbf{A} \mle - \boldsymbol{\beta} ) 
 \nonumber 
\end{align*}
As $ n \to \infty $, $ \mle \sim \mathcal{N}_{p} ( \textbf{0} , \boldsymbol{\mathcal{I}}^{-1} ) $. Therefore, $ \boldsymbol{\kappa}^n_1 \xrightarrow{ D } \boldsymbol{\kappa}_1 \sim \mathcal{N}_{p} \bigg( E( \boldsymbol{\kappa}_1 ) , Var ( \boldsymbol{\kappa}_1 ) \bigg) $ where
\begin{align*}
E( \boldsymbol{\kappa}_1 )  &  =  E [ \textbf{A} \,\mle  - \boldsymbol{\beta} ]
\nonumber  \\
& =  \textbf{A} \, \boldsymbol{\beta} - \boldsymbol{\beta}
 \nonumber  \\
 & = ( \textbf{A} \, - \textbf{I}_p ) \boldsymbol{\beta}
  \nonumber  
\end{align*}
\begin{align*}
Var( \boldsymbol{\kappa}_1 )  &  = Var [ \textbf{A} \,\mle  - \boldsymbol{\beta} ]
 \nonumber  \\
& = \textbf{A} \, Var( \mle ) \, \textbf{A}^\prime
 \nonumber  \\
 & = \textbf{A} \, \boldsymbol{\mathcal{I}}^{-1} \, \textbf{A}^\prime
\end{align*}

We show $ \boldsymbol{\kappa}^n_2 $ is asymptotically normally distributed with mean and variance as follow
\begin{align*}
 \boldsymbol{\kappa}^n_2 & = \sqrt{n} ( \mathbf{H} \ur - \mathbf{h} ) 
 \nonumber  \\
 & =  \sqrt{n} ( \mathbf{H}\, \ur - \mathbf{H}\, \boldsymbol{\beta} + \mathbf{H}\, \boldsymbol{\beta} - \mathbf{h} ) 
 \nonumber  \\
 & = \mathbf{H} \,  \sqrt{n} ( \ur - \boldsymbol{\beta} ) + \underbrace{ \sqrt{n} \,   ( \mathbf{H}\, \boldsymbol{\beta} - \mathbf{h} ) }_{ \boldsymbol{\vartheta}}
 \nonumber  \\
 & = \mathbf{H} \,  \boldsymbol{\kappa}^n_1  + \boldsymbol{\vartheta}
 \nonumber
\end{align*}
as $ n \to \infty $, $ \boldsymbol{\kappa}^n_1 \xrightarrow{ D } \boldsymbol{\kappa}_1 \sim \mathcal{N}_{p} \bigg ( ( \textbf{A} \, - \textbf{I}_p ) \boldsymbol{\beta} ,  \textbf{A} \, \boldsymbol{\mathcal{I}}^{-1} \, \textbf{A}^\prime \bigg ) $. So, $  \boldsymbol{\kappa}^n_2 \xrightarrow{ D }  \boldsymbol{\kappa}_2 \sim \mathcal{N}_{p} \bigg( E( \boldsymbol{\kappa}_2 ) , Var ( \boldsymbol{\kappa}_2 ) \bigg) $ where
\begin{align*}
E( \boldsymbol{\kappa}_2 )  &  =  \mathbf{H}\, E( \boldsymbol{\kappa}_1 ) + \boldsymbol{\vartheta} 
\nonumber  \\
& = \mathbf{H}\, ( \textbf{A} \, - \textbf{I}_p )\, \boldsymbol{\beta} + \boldsymbol{\vartheta} 
 \nonumber  
\end{align*}

\begin{align*}
Var( \boldsymbol{\kappa}_2 )  &  =  \mathbf{H}\, Var( \boldsymbol{\kappa}_1 )\,  \mathbf{H}^\prime
 \nonumber  \\
& =  \mathbf{H}\,  \textbf{A} \, \boldsymbol{\mathcal{I}}^{-1} \, \textbf{A}^\prime \mathbf{H}^\prime
 \nonumber
\end{align*}

Then 
\begin{align*}
\boldsymbol{\kappa}_3 & = \sqrt{n} ( \rr - \boldsymbol{\beta} ) 
 \nonumber  \\
 & =  \sqrt{n} ( \ur  -  \boldsymbol{\mathcal{J}} ( \textbf{H} \ur - \textbf{h}  ) - \boldsymbol{\beta} )
 \nonumber \\
 & = \sqrt{n} (  \ur - \boldsymbol{\beta} ) - \boldsymbol{\mathcal{J}}  \sqrt{n} ( \textbf{H}  \ur - \textbf{h}  ) 
 \nonumber  \\
 & =\boldsymbol{\kappa}^n_1 - \boldsymbol{\mathcal{J}}  \, \boldsymbol{\kappa}^n_2
 \nonumber  \\
 & = \boldsymbol{\kappa}^n_1- \boldsymbol{\mathcal{J}}  \, ( \mathbf{H} \, \boldsymbol{\kappa}^n_1  + \boldsymbol{\vartheta} )
 \nonumber  \\
 & = ( \mathbf{I}_p  - \boldsymbol{\mathcal{J}} \,  \mathbf{H} ) \, \boldsymbol{\kappa}^n_1 - \boldsymbol{\mathcal{J}}  \,\boldsymbol{\vartheta} ,
 \nonumber 
\end{align*}
As $ n \to \infty $, $ \boldsymbol{\kappa}^n_1 \xrightarrow{ D } \boldsymbol{\kappa}_1 \sim \mathcal{N}_{p} \bigg ( ( \textbf{A} \, - \textbf{I}_p ) \boldsymbol{\beta} ,  \textbf{A} \, \boldsymbol{\mathcal{I}}^{-1} \, \textbf{A}^\prime \bigg ) $. Thus, $ \boldsymbol{\kappa}^n_3 \xrightarrow{ D } \boldsymbol{\kappa}_3 \sim \mathcal{N}_{p} \bigg( E( \boldsymbol{\kappa}_3 ) , Var ( \boldsymbol{\kappa}_3 ) \bigg) $ where
\begin{align*}
E( \boldsymbol{\kappa}_3 )  &  =  E \big[ ( \mathbf{I}_p  - \boldsymbol{\mathcal{J}}  \,  \mathbf{H} ) \, \boldsymbol{\kappa}_1 - \boldsymbol{\mathcal{J}}  \, \boldsymbol{\vartheta} \big]
 \nonumber  \\
 & =  ( \mathbf{I}_p  - \boldsymbol{\mathcal{J}}  \,  \mathbf{H} ) \, E[ \boldsymbol{\kappa}_1 ] - \boldsymbol{\mathcal{J}} \,\boldsymbol{\vartheta} 
 \nonumber  \\
 & =  ( \mathbf{I}_p  - \boldsymbol{\mathcal{J}}  \,  \mathbf{H} ) \, ( \textbf{A} - \textbf{I}_p ) \, \boldsymbol{\beta} - \boldsymbol{\mathcal{J}} \,\boldsymbol{\vartheta} 
 \nonumber  
\end{align*}
\begin{align*}
Var ( \boldsymbol{\kappa}_3 ) & = Var \big [ ( \mathbf{I}_p  - \boldsymbol{\mathcal{J}}  \,  \mathbf{H} ) \, \boldsymbol{\kappa}_1 - \boldsymbol{\mathcal{J}}  \,\boldsymbol{\vartheta} \big]
 \nonumber  \\
 & = Var \big[ ( \mathbf{I}_p  - \boldsymbol{\mathcal{J}}  \,  \mathbf{H} ) \, \boldsymbol{\kappa}_1 \big]
 \nonumber  \\
 & = ( \mathbf{I}_p  - \boldsymbol{\mathcal{J}}  \,  \mathbf{H} ) \, \textbf{A} \, \boldsymbol{\mathcal{I}}^{-1} \, \textbf{A}^\prime  ( \mathbf{I}_p  - \boldsymbol{\mathcal{J}} \,  \mathbf{H} )^\prime
 \nonumber \\
 & = \textbf{A} \, \boldsymbol{\mathcal{I}}^{-1} \, \textbf{A}^\prime - \boldsymbol{\mathcal{J}} \,  \mathbf{H} \, \textbf{A} \, \boldsymbol{\mathcal{I}}^{-1} \, \textbf{A}^\prime
  \nonumber
\end{align*}


Now we write
\begin{align*}
\boldsymbol{\kappa}^n_4 & = \sqrt{n} ( \ur - \rr )
 \nonumber  \\
 & =  \sqrt{n}  \big ( \ur - \big[ \ur -  \boldsymbol{\mathcal{J}} ( \textbf{H} \ur - \textbf{h}  ) \big] \big)
 \nonumber \\
 & = \boldsymbol{\mathcal{J}} \,  \boldsymbol{\kappa}^n_2  
 \nonumber  \\
 & = \boldsymbol{\mathcal{J}} \, ( \mathbf{H} \,  \boldsymbol{\kappa}^n_1  + \boldsymbol{\vartheta} )
 \nonumber  \\
 & = \boldsymbol{\mathcal{J}} \,  \mathbf{H} \,  \boldsymbol{\kappa}^n_1  + \boldsymbol{\mathcal{J}} \, \boldsymbol{\vartheta} 
 \nonumber 
\end{align*}

As $ n \to \infty $, $ \boldsymbol{\kappa}^n_1 \xrightarrow{ D } \boldsymbol{\kappa}_1 \sim \mathcal{N}_{p} \bigg ( ( \textbf{A} \, - \textbf{I}_p ) \boldsymbol{\beta} ,  \textbf{A} \, \boldsymbol{\mathcal{I}}^{-1} \, \textbf{A}^\prime \bigg ) $. Thus, $ \boldsymbol{\kappa}^n_4 \xrightarrow{ D } \boldsymbol{\kappa}_4 \sim \mathcal{N}_{p} \bigg( E( \boldsymbol{\kappa}_4 ) , Var ( \boldsymbol{\kappa}_4 ) \bigg) $ where

\begin{align*}
E( \boldsymbol{\kappa}_4 )  &  = E\big[ \boldsymbol{\mathcal{J}} \,  \mathbf{H} \,  \boldsymbol{\kappa}_1  + \boldsymbol{\mathcal{J}} \, \boldsymbol{\vartheta} \big] 
 \nonumber \\
  &  = \ \boldsymbol{\mathcal{J}} \,  \mathbf{H} \,  E [ \boldsymbol{\kappa}_1 ] + \boldsymbol{\mathcal{J}} \, \boldsymbol{\vartheta} 
 \nonumber \\
 & = \boldsymbol{\mathcal{J}} \mathbf{H} \, (\textbf{A} - \mathbf{I}_p) \boldsymbol{\beta} + \boldsymbol{\mathcal{J}} \boldsymbol{\vartheta}
 \nonumber 
 \end{align*}
\begin{align*}
Var( \boldsymbol{\kappa}_4 )  &  = Var [ \boldsymbol{\mathcal{J}} \,  \mathbf{H} \,   \boldsymbol{\kappa}_1   + \boldsymbol{\mathcal{J}} \, \boldsymbol{\vartheta} ]
 \nonumber  \\
 & = ( \boldsymbol{\mathcal{J}} \,  \mathbf{H} ) \, Var [ \boldsymbol{\kappa}_1  ] \, ( \boldsymbol{\mathcal{J}} \,  \mathbf{H} )^\prime
 \nonumber  \\
 & =  \boldsymbol{\mathcal{J}} \,  \mathbf{H}  \,  \textbf{A} \, \boldsymbol{\mathcal{I}}^{-1} \, \textbf{A}^\prime  \, ( \boldsymbol{\mathcal{J}} \,  \mathbf{H} )^\prime
 \nonumber \\
 & =  \boldsymbol{\mathcal{J}} \,  \mathbf{H}  \,  \textbf{A} \, \boldsymbol{\mathcal{I}}^{-1} \, \textbf{A}^\prime .
 \nonumber
\end{align*}

The joint distribution of $ \boldsymbol{\kappa}^n_1 $ and $ \boldsymbol{\kappa}^n_4 $ is as follows:

\begin{align*}
\begin{pmatrix}
\boldsymbol{\kappa}^n_1 \\ \boldsymbol{\kappa}^n_4
\end{pmatrix}   & = 
\begin{pmatrix}
\mathbf{I}_p \,  \boldsymbol{\kappa}^n_1  + \textbf{0} \\ \boldsymbol{\mathcal{J}} \,  \mathbf{H} \, \boldsymbol{\kappa}^n_1  + \boldsymbol{\mathcal{J}} \, \boldsymbol{\vartheta} 
\end{pmatrix} 
 \nonumber \\
  &  = \begin{pmatrix}
\mathbf{I}_p  \\ \boldsymbol{\mathcal{J}} \,  \mathbf{H} 
\end{pmatrix}  \, \boldsymbol{\kappa}^n_1  + 
\begin{pmatrix}
\textbf{0}  \\ \boldsymbol{\mathcal{J}} \, \boldsymbol{\vartheta}  
\end{pmatrix} ,
 \nonumber 
\end{align*}

as $ n \to \infty $, $ \boldsymbol{\kappa}^n_1 \xrightarrow{ D } \boldsymbol{\kappa}_1 \sim \mathcal{N}_{p} \bigg ( ( \textbf{A} \, - \textbf{I}_p ) \boldsymbol{\beta} ,  \textbf{A} \, \boldsymbol{\mathcal{I}}^{-1} \, \textbf{A}^\prime \bigg ) $. So, $ \begin{pmatrix}
\boldsymbol{\kappa}^n_1  \\ \boldsymbol{\kappa}^n_4
\end{pmatrix} \xrightarrow{ D } \begin{pmatrix}
\boldsymbol{\kappa}_1   \\ \boldsymbol{\kappa}_4
\end{pmatrix} \sim \mathcal{N}_{2p} \bigg( \boldsymbol{\mu}_1 , \boldsymbol{\sigma}_1 \bigg) $ where

\begin{align*}
\boldsymbol{\mu}_1  & = E \Bigg [   
\begin{pmatrix}
\mathbf{I}_p  \\ \boldsymbol{\mathcal{J}} \,  \mathbf{H} 
\end{pmatrix}  \, \boldsymbol{\kappa}_1   + 
\begin{pmatrix}
\textbf{0}  \\ \boldsymbol{\mathcal{J}} \, \boldsymbol{\vartheta}  
\end{pmatrix}   
\Bigg ]
\nonumber \\
  & = 
\begin{pmatrix}
\mathbf{I}_p  \\ \boldsymbol{\mathcal{J}} \,  \mathbf{H} 
\end{pmatrix}  \, E [ \boldsymbol{\kappa}_1  ] + 
\begin{pmatrix}
\textbf{0}  \\ \boldsymbol{\mathcal{J}} \, \boldsymbol{\vartheta}  
\end{pmatrix}
 \nonumber \\
  &  = \begin{pmatrix}
( \textbf{A} \, - \textbf{I}_p ) \boldsymbol{\beta}  \\  \boldsymbol{\mathcal{J}} \,  \mathbf{H} ( \textbf{A} \, - \textbf{I}_p ) \boldsymbol{\beta} + \boldsymbol{\mathcal{J}} \, \boldsymbol{\vartheta}  
\end{pmatrix}
 \nonumber 
 \end{align*}
 
and 
 
\begin{align*}
\boldsymbol{\sigma}_1 &  = Var \Bigg [   
\begin{pmatrix}
\mathbf{I}_p  \\ \boldsymbol{\mathcal{J}} \,  \mathbf{H} 
\end{pmatrix}  \,  \boldsymbol{\kappa}_1 + 
\begin{pmatrix}
\textbf{0}  \\ \boldsymbol{\mathcal{J}} \, \boldsymbol{\vartheta}  
\end{pmatrix}   
\Bigg ]
\nonumber \\
 & = Var \Bigg [   
\begin{pmatrix}
\mathbf{I}_p  \\ \boldsymbol{\mathcal{J}} \,  \mathbf{H} 
\end{pmatrix}  \, \boldsymbol{\kappa}_1
\Bigg ]
\nonumber \\
  & = 
\begin{pmatrix}
\mathbf{I}_p  \\ \boldsymbol{\mathcal{J}} \,  \mathbf{H} 
\end{pmatrix} \, Var [ \, \boldsymbol{\kappa}_1 ]\, 
 \begin{pmatrix}
\mathbf{I}_p &  ( \boldsymbol{\mathcal{J}} \,  \mathbf{H} )^\prime
\end{pmatrix}
\nonumber \\
   & = 
\begin{pmatrix}
\mathbf{I}_p  \\ \boldsymbol{\mathcal{J}} \,  \mathbf{H} 
\end{pmatrix} \,  \textbf{A} \, \boldsymbol{\mathcal{I}}^{-1} \, \textbf{A}^\prime  \, 
 \begin{pmatrix}
\mathbf{I}_p  &  ( \boldsymbol{\mathcal{J}} \,  \mathbf{H} )^\prime
\end{pmatrix}
 \nonumber \\
  & = 
\begin{pmatrix}
\textbf{A} \, \boldsymbol{\mathcal{I}}^{-1} \, \textbf{A}^\prime & \textbf{A} \, \boldsymbol{\mathcal{I}}^{-1} \, \textbf{A}^\prime \, ( \boldsymbol{\mathcal{J}} \,  \mathbf{H} )^\prime \\ 
\boldsymbol{\mathcal{J}} \,  \mathbf{H} \, \textbf{A} \, \boldsymbol{\mathcal{I}}^{-1} \, \textbf{A}^\prime & \boldsymbol{\mathcal{J}} \,  \mathbf{H} \, \textbf{A} \, \boldsymbol{\mathcal{I}}^{-1} \, \textbf{A}^\prime \,  ( \boldsymbol{\mathcal{J}} \,  \mathbf{H} )^\prime
\end{pmatrix}
  \nonumber \\
  & = 
\begin{pmatrix}
\textbf{A} \, \boldsymbol{\mathcal{I}}^{-1} \, \textbf{A}^\prime & \boldsymbol{\mathcal{J}} \,  \mathbf{H} \, \textbf{A} \, \boldsymbol{\mathcal{I}}^{-1} \, \textbf{A}^\prime   \\ 
\boldsymbol{\mathcal{J}} \,  \mathbf{H} \, \textbf{A} \, \boldsymbol{\mathcal{I}}^{-1} \, \textbf{A}^\prime  & \boldsymbol{\mathcal{J}} \,  \mathbf{H} \, \textbf{A} \, \boldsymbol{\mathcal{I}}^{-1} \, \textbf{A}^\prime 
\end{pmatrix}
 \nonumber 
\end{align*}


In a similar way, we consider

\begin{align*}
\begin{pmatrix}
\boldsymbol{\kappa}^n_3 \\ \boldsymbol{\kappa}^n_4
\end{pmatrix}   & = 
\begin{pmatrix}
( \mathbf{I}_p -  \boldsymbol{\mathcal{J}} \,  \mathbf{H} )\, \boldsymbol{\kappa}_1  - \boldsymbol{\mathcal{J}} \, \boldsymbol{\vartheta} \\
\boldsymbol{\mathcal{J}} \,  \mathbf{H}  \, \boldsymbol{\kappa}_1 +  \boldsymbol{\mathcal{J}} \, \boldsymbol{\vartheta}
\end{pmatrix} 
 \nonumber \\
  &  = \begin{pmatrix}
\mathbf{I}_p -  \boldsymbol{\mathcal{J}} \,  \mathbf{H} \\
\boldsymbol{\mathcal{J}} \,  \mathbf{H}
\end{pmatrix}  \, \boldsymbol{\kappa}_1  + 
\begin{pmatrix}
- \mathbf{I}_p   \\ \mathbf{I}_p  
\end{pmatrix} \, \boldsymbol{\mathcal{J}} \, \boldsymbol{\vartheta}
 \nonumber 
\end{align*}

Therefore, as $ n \to \infty $, $ \boldsymbol{\kappa}^n_1 \xrightarrow{ D } \boldsymbol{\kappa}_1 \sim \mathcal{N}_{p} \bigg ( ( \textbf{A} \, - \textbf{I}_p ) \boldsymbol{\beta} ,  \textbf{A} \, \boldsymbol{\mathcal{I}}^{-1} \, \textbf{A}^\prime \bigg ) $. So, $ \begin{pmatrix}
\boldsymbol{\kappa}^n_3 \\ \boldsymbol{\kappa}^n_4
\end{pmatrix} \xrightarrow{ D } \begin{pmatrix}
\boldsymbol{\kappa}_3 \\ \boldsymbol{\kappa}_4
\end{pmatrix} \sim \mathcal{N}_{2p} \bigg( \boldsymbol{\mu}_2 , \boldsymbol{\sigma}_2 \bigg) $ where

\begin{align*}
\boldsymbol{\mu}_2  & = E \Bigg [   
\begin{pmatrix}
\mathbf{I}_p -  \boldsymbol{\mathcal{J}} \,  \mathbf{H} \\
\boldsymbol{\mathcal{J}} \,  \mathbf{H}
\end{pmatrix}  \, \boldsymbol{\kappa}_1  + 
\begin{pmatrix}
- \mathbf{I}_p   \\ \mathbf{I}_p  
\end{pmatrix} \, \boldsymbol{\mathcal{J}} \, \boldsymbol{\vartheta}  
\Bigg ]
\nonumber \\
  & = 
\begin{pmatrix}
\mathbf{I}_p -  \boldsymbol{\mathcal{J}} \,  \mathbf{H} \\
\boldsymbol{\mathcal{J}} \,  \mathbf{H}
\end{pmatrix}  \, E [ \boldsymbol{\kappa}_1 ] + 
\begin{pmatrix}
- \mathbf{I}_p   \\ \mathbf{I}_p  
\end{pmatrix} \, \boldsymbol{\mathcal{J}} \, \boldsymbol{\vartheta} 
 \nonumber \\
  &  = \begin{pmatrix}
( \mathbf{I}_p -  \boldsymbol{\mathcal{J}} \,  \mathbf{H} )\, ( \textbf{A} - \mathbf{I}_p  ) \,\boldsymbol{\beta} - \boldsymbol{\mathcal{J}} \, \boldsymbol{\vartheta}   \\  \boldsymbol{\mathcal{J}} \,  \mathbf{H} \, ( \textbf{A} - \mathbf{I}_p  ) \,\boldsymbol{\beta} +  \boldsymbol{\mathcal{J}} \, \boldsymbol{\vartheta}  
\end{pmatrix} 
 \nonumber 
 \end{align*}
 
 and 
 
\begin{align*}
\boldsymbol{\sigma}_2 &  = Vae \Bigg [   
\begin{pmatrix}
\mathbf{I}_p -  \boldsymbol{\mathcal{J}} \,  \mathbf{H} \\
\boldsymbol{\mathcal{J}} \,  \mathbf{H}
\end{pmatrix}  \, \boldsymbol{\kappa}_1  + 
\begin{pmatrix}
- \mathbf{I}_p   \\ \mathbf{I}_p 
\end{pmatrix} \, \boldsymbol{\mathcal{J}} \, \boldsymbol{\vartheta}  
\Bigg ]
\nonumber \\
 & = \begin{pmatrix}
\mathbf{I}_p -  \boldsymbol{\mathcal{J}} \,  \mathbf{H} \\
\boldsymbol{\mathcal{J}} \,  \mathbf{H}
\end{pmatrix}  \, Var [ \boldsymbol{\kappa}_1  ] \, 
\begin{pmatrix}
(\mathbf{I}_p -  \boldsymbol{\mathcal{J}} \,  \mathbf{H} )^\prime & 
(\boldsymbol{\mathcal{J}} \,  \mathbf{H} )^\prime
\end{pmatrix}
\nonumber \\
  & = 
\begin{pmatrix}
\mathbf{I}_p -  \boldsymbol{\mathcal{J}} \,  \mathbf{H} \\
\boldsymbol{\mathcal{J}} \,  \mathbf{H}
\end{pmatrix}  \, \textbf{A} \, \boldsymbol{\mathcal{I}}^{-1} \, \textbf{A}^\prime \, 
\begin{pmatrix}
(\mathbf{I}_p-  \boldsymbol{\mathcal{J}} \,  \mathbf{H} )^\prime & 
(\boldsymbol{\mathcal{J}} \,  \mathbf{H} )^\prime
\end{pmatrix}
\nonumber \\
   & = 
\begin{pmatrix}
(\mathbf{I}_p -  \boldsymbol{\mathcal{J}} \,  \mathbf{H} ) \, \textbf{A} \, \boldsymbol{\mathcal{I}}^{-1} \, \textbf{A}^\prime  \, (\mathbf{I}_p -  \boldsymbol{\mathcal{J}} \,  \mathbf{H} )^\prime  &  
(\mathbf{I}_p-  \boldsymbol{\mathcal{J}} \,  \mathbf{H} ) \, \textbf{A} \, \boldsymbol{\mathcal{I}}^{-1} \, \textbf{A}^\prime  \, (\boldsymbol{\mathcal{J}} \,  \mathbf{H} )^\prime  \\
\boldsymbol{\mathcal{J}} \,  \mathbf{H} \, \textbf{A} \, \boldsymbol{\mathcal{I}}^{-1} \, \textbf{A}^\prime  \, (\mathbf{I}_p -  \boldsymbol{\mathcal{J}} \,  \mathbf{H} )^\prime  &
\boldsymbol{\mathcal{J}} \,  \mathbf{H} \, \textbf{A} \, \boldsymbol{\mathcal{I}}^{-1} \, \textbf{A}^\prime \,( \boldsymbol{\mathcal{J}} \,  \mathbf{H} )^\prime 
\end{pmatrix}
 \nonumber \\
  & = 
\begin{pmatrix}
\textbf{A} \, \boldsymbol{\mathcal{I}}^{-1} \, \textbf{A}^\prime -  \boldsymbol{\mathcal{J}} \,  \mathbf{H} \, \textbf{A} \, \boldsymbol{\mathcal{I}}^{-1} \, \textbf{A}^\prime & \textbf{0} \\
\textbf{0} &   \boldsymbol{\mathcal{J}} \,  \mathbf{H} \, \textbf{A} \, \boldsymbol{\mathcal{I}}^{-1} \, \textbf{A}^\prime 
\end{pmatrix}.
 \nonumber 
\end{align*}


\newpage
\section*{Appendix 3. Proof of Theorem 3.2}

Here, we provide the proof of bias expressions. Based on Lemma 3.1 we have
\begin{equation*}
\mathcal{B} ( \ur )   = \lim_{ n \to \infty } \, E \bigg[ \sqrt{n} \,\bigg ( \ur - \boldsymbol{\beta} \bigg) \bigg] = E [ \boldsymbol{\kappa}_1 ] = ( \textbf{A} - \textbf{I}_p )\, \boldsymbol{\beta}
\end{equation*}

\begin{align*}
\mathcal{B} ( \rr )  & = \lim_{ n \to \infty } \, E \bigg[ \sqrt{n} \,\bigg ( \rr - \boldsymbol{\beta} \bigg) \bigg] 
\nonumber \\
& = E [ \boldsymbol{\kappa}_3 ] 
\nonumber \\
& ( \textbf{I}_p - \boldsymbol{\mathcal{J}} \,  \mathbf{H} ) ( \textbf{A} - \textbf{I}_p ) \,\boldsymbol{\beta}  - \boldsymbol{\mathcal{J}} \,   \boldsymbol{\vartheta}  
\end{align*}

\begin{align*}
\mathcal{B} ( \rls )  & = \lim_{ n \to \infty } \, E \bigg[ \sqrt{n} \, \bigg( \rls - \boldsymbol{\beta} \bigg) \bigg] 
 \nonumber \\
   & =  \lim_{ n \to \infty } \, E \bigg[ \sqrt{n}\, \bigg( \delta \, \rr + ( 1 - \delta ) \, \ur    - \boldsymbol{\beta} \bigg) \bigg] 
 \nonumber \\
  & =  \lim_{ n \to \infty } \, E \bigg[ \sqrt{n}\, \bigg( \ur  - \boldsymbol{\beta} \bigg) - \delta \, \sqrt{n} \bigg ( \ur  -  \rr \bigg) \bigg]
  \nonumber \\
  & = E (  \boldsymbol{\kappa}_1  ) - \delta \, E ( \boldsymbol{\kappa}_4 )
  \nonumber \\
  & = \mathcal{B} ( \ur )  - \delta \,\boldsymbol{\mathcal{J}}\, [ \textbf{H}   ( \textbf{A} - \textbf{I}_p ) \,\boldsymbol{\beta} + \boldsymbol{\vartheta} ]
   \nonumber 
\end{align*}

\begin{align*}
\mathcal{B} ( \spe )   & =   \lim_{ n \to \infty } \, E \bigg[ \sqrt{n}\, \bigg( \spe - \boldsymbol{\beta} \bigg) \bigg] 
 \nonumber \\
 & = \lim_{ n \to \infty } \, E \bigg[ \sqrt{n}\, \bigg( \ur - \delta \, ( \ur - \rr ) \, I ( T_n \leq T_{n,\alpha} ) - \boldsymbol{\beta} \bigg) \bigg]
 \nonumber \\
 & = \lim_{ n \to \infty } \, E \bigg [ \sqrt{n} \bigg( \ur - \boldsymbol{\beta} \bigg) - \delta \, \sqrt{n}\, \bigg( \ur - \rr \bigg) \, I ( T_n \leq T_{n,\alpha} )  \bigg]
 \nonumber \\
 & = E [ \boldsymbol{\kappa}_1 ] - \delta \, E \bigg[ \boldsymbol{\kappa}_4 \, I ( T_n \leq T_{n,\alpha} ) \bigg],
 \nonumber
\end{align*}
based on Equation (\ref{equ21}) we can write
\begin{align*}
\mathcal{B} ( \spe )   & = E [ \boldsymbol{\kappa}_1 ]   - \delta \, E [ \boldsymbol{\kappa}_4 ]   \, P ( \chi^2_{p_2 + 2} ( \Delta^{*} ) \leq \chi^2_{p_2, \alpha} )
\nonumber \\
 & = \mathcal{B} ( \ur ) -  \delta \,\boldsymbol{\mathcal{J}}\, [ \textbf{H}   ( \textbf{A} - \textbf{I}_p ) \,\boldsymbol{\beta} + \boldsymbol{\vartheta} ] \, P ( \chi^2_{p_2 + 2} ( \Delta^{*} ) \leq \chi^2_{p_2, \alpha} )
\nonumber \\
 & = \mathcal{B} ( \ur ) -  \delta \,\boldsymbol{\mathcal{J}}\, [ \textbf{H}   ( \textbf{A} - \textbf{I}_p ) \,\boldsymbol{\beta} + \boldsymbol{\vartheta} ] \,\boldsymbol{ \Psi}_{p_2 + 2} ( \chi^2_{p_2, \alpha} ; \Delta^{*}) .
\end{align*}
If $ \delta = 1 $,
\begin{equation*}
\mathcal{B} ( \rpt ) =  \mathcal{B} ( \ur ) -  \boldsymbol{\mathcal{J}}\, [ \textbf{H}   ( \textbf{A} - \textbf{I}_p ) \,\boldsymbol{\beta} + \boldsymbol{\vartheta} ]\,\boldsymbol{ \Psi}_{p_2 + 2} ( \chi^2_{p_2, \alpha} ; \Delta^{*}) .
\end{equation*}


In a similar way, we can obtain 
\begin{align*}
\mathcal{B} ( \rs )   & =   \lim_{ n \to \infty } \, E \bigg[ \sqrt{n}\, \bigg( \rs - \boldsymbol{\beta} \bigg) \bigg] 
\nonumber \\
&  =  \lim_{ n \to \infty } \, E \bigg[ \sqrt{n} \, \bigg( \rr + ( 1 - ( p_2 - 2 ) \, T_n^{-1} ) \, ( \ur - \rr ) - \boldsymbol{\beta}  \bigg) \bigg]  
\nonumber \\
& = \lim_{ n \to \infty } \,\Bigg \lbrace E \bigg[ \sqrt{n} \, \bigg( \ur -  \boldsymbol{\beta} \bigg) \bigg] - ( p_2 - 2 ) \,E\bigg [ T_n^{-1} \, \sqrt{n} \, \bigg ( \ur - \rr \bigg) \bigg] \Bigg \rbrace
\nonumber \\
& = E [ \boldsymbol{\kappa}_1 ] - ( p_2 - 2 ) \, E [ T_n^{-1}\,  \boldsymbol{\kappa}_4  ]
\nonumber \\
& = \mathcal{B} ( \ur ) - ( p_2 - 2 ) \,E [ \boldsymbol{\kappa}_4 ]\, E \bigg [ \frac{ 1 }{ \chi^2_{ p_2 + 2 } ( \Delta^{*} ) } \bigg ]
\nonumber \\
& = \mathcal{B} ( \ur ) - ( p_2 - 2 ) \, \boldsymbol{\mathcal{J}} \ [ \textbf{H} \,( \textbf{A} - \textbf{I}_p )\, \boldsymbol{\beta}   + \boldsymbol{\vartheta} ] \, E  \Big[  \frac{1}{\chi^2_{ p_2 + 2 } ( \Delta^{*} )}  \Big].
\nonumber 
\end{align*}

Also,
\begin{align*} 
\mathcal{B} ( \rps ) & = \lim_{ n \to \infty } \, E \bigg[ \sqrt{n}\, \bigg( \rps - \boldsymbol{\beta} \bigg) \bigg] 
\nonumber \\
& = \lim_{ n \to \infty } \, E \bigg [ \sqrt{n} \,\bigg ( \mathcal{B} ( \rs  - \boldsymbol{\beta} \bigg ) - ( 1 - (p_2 - 2) \, T_n^{-1} )  
\nonumber \\
& \qquad \qquad \qquad \qquad \times \sqrt{n} \,\bigg ( ( \ur - \rr )  \, I ( T_n < p_2 - 2 ) \bigg ) \bigg]
\nonumber \\
& = \mathcal{B} ( \rs ) - \lim_{ n \to \infty } \, E \bigg [ \sqrt{n}\, \bigg ( \ur - \rr \bigg ) \, ( 1 - (p_2 - 2) \, T_n^{-1} ) \, I ( T_n < p_2 - 2 ) \bigg ] 
\nonumber \\
& = \mathcal{B} ( \rs ) - E [ \boldsymbol{\kappa}_4 \,( 1 - (p_2 - 2) \, T_n^{-1} ) \,I ( T_n < p_2 - 2 ) ]
\nonumber \\
& = \mathcal{B} ( \rs ) - E [ \boldsymbol{\kappa}_4 \,I ( T_n < p_2 - 2 ) ] + (p_2 - 2)\, E [ \boldsymbol{\kappa}_4 \, T_n^{-1} \,I ( T_n < p_2 - 2 )] 
\nonumber \\
& = \mathcal{B} ( \rs ) - E [ \boldsymbol{\kappa}_4 ] \, \boldsymbol{\Psi}_{p_2 + 2}(\chi^2_{p_2 ,\alpha} ; \Delta^{*} ) + (p_2 - 2)\, E [ \boldsymbol{\kappa}_4 ] E \bigg [ \frac{ I ( T_n < p_2 - 2 ) }{ \chi^2_{p_2 + 2} ( \Delta^{*} ) } \bigg ] 
\nonumber \\ 
& = \mathcal{B} ( \rs ) - E [ \boldsymbol{\kappa}_4 ] \, \bigg \lbrace  \boldsymbol{\Psi}_{p_2 + 2}(\chi^2_{p_2 ,\alpha} ; \Delta^{*} ) + (p_2 - 2)\, E \bigg [ \frac{ I ( T_n < p_2 - 2 ) }{ \chi^2_{p_2 + 2} ( \Delta^{*} ) } \bigg ] \bigg \rbrace
\nonumber \\ 
& = \mathcal{B} ( \rs ) - \, \boldsymbol{\mathcal{J}}\, [ \textbf{H} (  \textbf{A} - \textbf{I}_p ) \,\boldsymbol{\beta} + \boldsymbol{\vartheta} ] 
\bigg \lbrace  \boldsymbol{\Psi}_{p_2 + 2}(\chi^2_{p_2 ,\alpha} ; \Delta^{*} ) 
\nonumber \\
& \qquad \qquad \qquad \qquad
+ ( p_2 - 2 ) \, E \biggl [ \dfrac{ I ( \chi^2_{p_2 + 2} (\Delta^{*}) < p_2 - 2 ) }{ \chi^2_{p_2 + 2} ( \Delta^{*} ) } \biggl ] \bigg \rbrace.
\nonumber
\end{align*}

\newpage
\section*{Appendix 4. Proof of Theorem 3.3}

Here, we compute the asymptotic covariance matrices of the estimators
\begin{align*}
\mathcal{V} ( \ur )  & = \lim_{n \to \infty} \, E \Big (  \sqrt{n} ( \ur - \boldsymbol{\beta} )   \,    \sqrt{n} ( \ur - \boldsymbol{\beta} )^\prime   \Big )  
\nonumber \\
& =  \lim_{n \to \infty} \, E ( \boldsymbol{\kappa}^n_1 \, \boldsymbol{\kappa}^n_1{^\prime} )
\nonumber \\
& =  E ( \boldsymbol{\kappa}_1 \, \boldsymbol{\kappa}_1^\prime )
\nonumber \\
& =  Var ( \boldsymbol{\kappa}_1 ) +  E ( \boldsymbol{\kappa}_1 \, E ( \boldsymbol{\kappa}_1{^\prime} )
\nonumber \\
& =  \textbf{A} \boldsymbol{\mathcal{I}}^{-1} \textbf{A}^\prime + \bigg[ ( \textbf{A} - \mathbf{I}_p ) \,\boldsymbol{\beta} \bigg] \, \bigg[ ( \textbf{A} - \mathbf{I}_p ) \, \boldsymbol{\beta} \bigg]^\prime
\nonumber 
\end{align*}
\begin{align*}
\mathcal{V} ( \rr )  & = \lim_{n \to \infty} \, E \Big (  \sqrt{n} ( \rr - \boldsymbol{\beta}   )\,    \sqrt{n} ( \rr - \boldsymbol{\beta} )^\prime   \Big )  
\nonumber \\
& =  \lim_{n \to \infty} \, E ( \boldsymbol{\kappa}^n_3 \, \boldsymbol{\kappa}^n_3{^\prime} )
\nonumber \\
& =  E ( \boldsymbol{\kappa}_3 \, \boldsymbol{\kappa}_3{^\prime} )
\nonumber \\
& =  Var ( \boldsymbol{\kappa}_3 ) +  E ( \boldsymbol{\kappa}_3 ) \, E ( \boldsymbol{\kappa}_3{^\prime} )
\nonumber \\
& = \textbf{A} \, \boldsymbol{\mathcal{I}}^{-1} \, \textbf{A}^\prime - \boldsymbol{\mathcal{J}} \,  \mathbf{H} \, \textbf{A} \, \boldsymbol{\mathcal{I}}^{-1} \, \textbf{A}^\prime  
\nonumber \\
& \qquad + \bigg[  ( \mathbf{I}_p  - \boldsymbol{\mathcal{J}}  \,  \mathbf{H} ) \, ( \textbf{A} - \textbf{I}_p ) \, \boldsymbol{\beta} - \boldsymbol{\mathcal{J}} \,\boldsymbol{\vartheta}  \bigg] \, \bigg [  ( \mathbf{I}_p  - \boldsymbol{\mathcal{J}}  \,  \mathbf{H} ) \, ( \textbf{A} - \textbf{I}_p ) \, \boldsymbol{\beta} - \boldsymbol{\mathcal{J}} \,\boldsymbol{\vartheta} \bigg ]^\prime
\nonumber 
\end{align*}

\begin{align*}
\mathcal{V} ( \rls )  & = \lim_{n \to \infty} \, E \Big (  \sqrt{n} ( \rls - \boldsymbol{\beta} )   \,   \sqrt{n} ( \rls - \boldsymbol{\delta} )^\prime   \Big )  
\nonumber \\
& =  \lim_{n \to \infty} \, E [ ( \boldsymbol{\kappa}^n_1 - \delta \, \boldsymbol{\kappa}^n_4  ) \, ( \boldsymbol{\kappa}^n_1 - \delta \, \boldsymbol{\kappa}^n_4  )^\prime ]
\nonumber \\
& = E [ ( \boldsymbol{\kappa}_1 - \delta \, \boldsymbol{\kappa}_4  ) \, ( \boldsymbol{\kappa}_1 - \delta \, \boldsymbol{\kappa}_4  )^\prime ]
\nonumber \\
& = E [ \boldsymbol{\kappa}_1 \, \boldsymbol{\kappa}_1^\prime  ] - 2 \delta \, E [ \boldsymbol{\kappa}_1 \, \boldsymbol{\kappa}_4^\prime  ] + \delta^2\, E [ \boldsymbol{\kappa}_4 \, \boldsymbol{\kappa}_4^\prime  ]
\nonumber \\
& = \mathcal{V} ( \ur )  -  2 \delta \, \underbrace{ E [ \boldsymbol{\kappa}_1 \, \boldsymbol{\kappa}_4^\prime  ] }_{e_1} + \delta^2\,  E [ \boldsymbol{\kappa}_4 \, \boldsymbol{\kappa}_4^\prime  ]
\nonumber 
\end{align*}
Using the conditional expectation of a multivariate normal distribution, $ e_1 $ becomes
\begin{align*}
e_1 & = E [ \boldsymbol{\kappa}_1 \, \boldsymbol{\kappa}_4^\prime ]
\nonumber \\
& = E \big (\, \boldsymbol{\kappa}_4^\prime \, E [ \boldsymbol{\kappa}_1  \, | \, \boldsymbol{\kappa}_4 ]   \big )
\nonumber \\
& = E \big ( \boldsymbol{\kappa}_4^\prime \big \lbrace  ( \textbf{A} - \textbf{I}_p ) \, \boldsymbol{\beta} +  \boldsymbol{\kappa}_4 - 
\boldsymbol{\mathcal{J}} \, \textbf{H}\, ( \textbf{A} - \textbf{I}_p ) \, \boldsymbol{\beta} -  \boldsymbol{\mathcal{J}} \, \boldsymbol{\vartheta}  \big \rbrace \big ) 
\nonumber \\
& = ( \textbf{A} - \textbf{I}_p ) \, \boldsymbol{\beta} \, E( \boldsymbol{\kappa}_4^\prime ) + E ( \boldsymbol{\kappa}_4\, \boldsymbol{\kappa}_4^\prime ) - [ \boldsymbol{\mathcal{J}}\, \textbf{H}\,( \textbf{A} - \textbf{I}_p ) \, \boldsymbol{\beta} + \boldsymbol{\mathcal{J}} \, \boldsymbol{\vartheta}  ]\, E( \boldsymbol{\kappa}_4^\prime )
\nonumber \\
& = E ( \boldsymbol{\kappa}_4\, \boldsymbol{\kappa}_4^\prime ) +  E( \boldsymbol{\kappa}_4^\prime ) \big \lbrace ( \textbf{I}_p - \boldsymbol{\mathcal{J}}\, \textbf{H} ) \, ( \textbf{A} - \textbf{I}_p ) \, \boldsymbol{\beta} ) - \boldsymbol{\mathcal{J}} \, \boldsymbol{\vartheta}
\big \rbrace
\nonumber 
\end{align*}
Therefore,
\begin{align*}
\mathcal{V} ( \rls ) &  = \mathcal{V} ( \ur )  
+ \delta^2\,  E [ \boldsymbol{\kappa}_4 \, \boldsymbol{\kappa}_4^\prime  ]
\nonumber \\
& -  2 \delta \, \bigg ( E ( \boldsymbol{\kappa}_4\, \boldsymbol{\kappa}_4^\prime )  + E( \boldsymbol{\kappa}_4^\prime ) \big \lbrace ( \textbf{I}_p - \boldsymbol{\mathcal{J}}\, \textbf{H} ) \, ( \textbf{A} - \textbf{I}_p ) \, \boldsymbol{\beta} ) - \boldsymbol{\mathcal{J}} \, \boldsymbol{\vartheta}  \bigg ) 
\nonumber \\
&  = \mathcal{V} ( \ur )  -  2 \delta  \big \lbrace ( \textbf{I}_p - \boldsymbol{\mathcal{J}}\, \textbf{H} ) \, ( \textbf{A} - \textbf{I}_p ) \, \boldsymbol{\beta} ) - \boldsymbol{\mathcal{J}} \, \boldsymbol{\vartheta} \big \rbrace  E( \boldsymbol{\kappa}_4^\prime )
- \delta \, ( 2 - \delta ) \,E [ \boldsymbol{\kappa}_4 \, \boldsymbol{\kappa}_4^\prime  ]
\nonumber \\
& = \mathcal{V} ( \ur )  - 2 \delta \, \big \lbrace ( \textbf{I}_p - \boldsymbol{\mathcal{J}}\, \textbf{H} ) \, ( \textbf{A} - \textbf{I}_p ) \, \boldsymbol{\beta} ) - \boldsymbol{\mathcal{J}} \, \boldsymbol{\vartheta} \big \rbrace
\big \lbrace  \boldsymbol{\mathcal{J}}\, \textbf{H}\,( \textbf{A} - \textbf{I}_p ) \, \boldsymbol{\beta} + \boldsymbol{\mathcal{J}} \, \boldsymbol{\vartheta} \big \rbrace ^\prime
\nonumber \\
& - \delta \, ( 2 - \delta ) \, \big \lbrace   \boldsymbol{\mathcal{J}}\, \textbf{H}\, \textbf{A}  \,  \boldsymbol{\mathcal{I}}^{-1} \, \textbf{A}^\prime +  [ \boldsymbol{\mathcal{J}}\, \textbf{H}\,( \textbf{A} - \textbf{I}_p ) \, \boldsymbol{\beta} + \boldsymbol{\mathcal{J}} \, \boldsymbol{\vartheta}  ] \, [ \boldsymbol{\mathcal{J}}\, \textbf{H}\,( \textbf{A} - \textbf{I}_p ) \, \boldsymbol{\beta} + \boldsymbol{\mathcal{J}} \, \boldsymbol{\vartheta}  ]^\prime \big \rbrace
\nonumber 
\end{align*}


Next we obtain $ \mathcal{V} ( \spe ) $ as follows
\begin{align*}
\mathcal{V} ( \spe ) & = \lim_{n \to \infty} \, E \Big (  \sqrt{n} ( \spe - \boldsymbol{\beta} )   \,  \sqrt{n} ( \spe - \boldsymbol{\beta} )^\prime   \Big )  
\nonumber \\
& =  \lim_{n \to \infty} \, E [\,  \lbrace \boldsymbol{\kappa}^n_1 - \delta \, \boldsymbol{\kappa}^n_4\, I ( T_n \leq  T_{n,\alpha} )  \rbrace \, \lbrace  \boldsymbol{\kappa}^n_1 - \delta \, \boldsymbol{\kappa}^n_4\, I ( T_n \leq  T_{n,\alpha} ) \rbrace^\prime  \, ]
\nonumber \\
& = \lim_{n \to \infty} \, E ( \boldsymbol{\kappa}^n_1 \, \boldsymbol{\kappa}^n_1{^\prime}  ) - 2 \delta \, \lim_{n \to \infty} \, E ( \boldsymbol{\kappa}^n_1 \, \boldsymbol{\kappa}^n_4{^\prime} \, I ( T_n \leq  T_{n,\alpha} ) ) 
\nonumber \\
& \qquad \qquad \qquad \qquad + \delta^2 \, \lim_{n \to \infty} \, E ( \boldsymbol{\kappa}^n_4 \, \boldsymbol{\kappa}^n_4{^\prime} \, I ( T_n \leq  T_{n,\alpha} ) ) 
\nonumber \\
& =  E ( \boldsymbol{\kappa}_1 \, \boldsymbol{\kappa}_1^\prime  ) - 2 \delta \, E ( \boldsymbol{\kappa}_1 \, \boldsymbol{\kappa}_4^\prime \, I ( \chi^2_{p_2} ( \Delta^{*} ) \leq  \chi^2_{p_2 , \alpha} ) ) + \delta^2  \, E ( \boldsymbol{\kappa}_4 \, \boldsymbol{\kappa}_4^\prime \, I ( \chi^2_{p_2} ( \Delta^{*} ) \leq  \chi^2_{p_2 , \alpha} ) ) 
\nonumber\\
& = \mathcal{V} ( \ur )  - 2 \delta \,\underbrace{  E ( \boldsymbol{\kappa}_1 \, \boldsymbol{\kappa}_4^\prime \, I ( \chi^2_{p_2} ( \Delta^{*} ) \leq  \chi^2_{p_2 , \alpha} ) ) }_{e_2} + \delta^2  \, E ( \boldsymbol{\kappa}_4 \, \boldsymbol{\kappa}_4^\prime \, I ( \chi^2_{p_2} ( \Delta^{*} ) \leq  \chi^2_{p_2 , \alpha} ) )   ,
\nonumber 
\end{align*}
by using conditional expectation, $ e_2 $ becomes
\begin{align*}
e_2 & =  E ( \boldsymbol{\kappa}_1 \, \boldsymbol{\kappa}_4^\prime \, I ( \chi^2_{p_2} ( \Delta^{*} ) \leq  \chi^2_{p_2 , \alpha} ) ) 
\nonumber\\
 & = E \big [ E ( \boldsymbol{\kappa}_1  \, \boldsymbol{\kappa}_4^\prime  \, I ( \chi^2_{p_2} ( \Delta^{*} ) \leq  \chi^2_{p_2 , \alpha} )   \, | \,  \boldsymbol{\kappa}_4 )  \big ]
\nonumber\\
 & = E \big [ E ( \boldsymbol{\kappa}_1 \, |\, \boldsymbol{\kappa}_4 )\, \boldsymbol{\kappa}_4 ^\prime \, I ( \chi^2_{p_2} ( \Delta^{*} ) \leq  \chi^2_{p_2 , \alpha} )  \big ]
\nonumber\\
 & = E \big [ \, \lbrace  ( \textbf{A} - \textbf{I}_p ) \, \boldsymbol{\beta} + \boldsymbol{\kappa}_4  -  \boldsymbol{\mathcal{J}} \, \textbf{H} \, ( \textbf{A} - \textbf{I}_p ) \, \boldsymbol{\beta} - \boldsymbol{\mathcal{J}} \, \boldsymbol{\vartheta} \rbrace  \, \boldsymbol{\kappa}_4 ^\prime 
\, I ( \chi^2_{p_2} ( \Delta^{*} ) \leq  \chi^2_{p_2 , \alpha} )  \big]
\nonumber\\
 & = ( \textbf{A} - \textbf{I}_p ) \, \boldsymbol{\beta} \, E [ \boldsymbol{\kappa}_4 ^\prime \, I ( \chi^2_{p_2} ( \Delta^{*} ) \leq  \chi^2_{p_2 , \alpha} ) ] + E [ \boldsymbol{\kappa}_4 \, \boldsymbol{\kappa}_4 ^\prime \, I ( \chi^2_{p_2} ( \Delta^{*} ) \leq  \chi^2_{p_2 , \alpha} ) ] 
 \nonumber\\
& - \big ( \boldsymbol{\mathcal{J}} \, \textbf{H} \, ( \textbf{A} - \textbf{I}_p ) \, \boldsymbol{\beta} + \boldsymbol{\mathcal{J}} \, \boldsymbol{\vartheta} \big ) \, E [ \boldsymbol{\kappa}_4 ^\prime \, I ( \chi^2_{p_2} ( \Delta^{*} ) \leq  \chi^2_{p_2 , \alpha} ) ]
\nonumber\\
& = \bigg \lbrace ( \textbf{I}_p - \boldsymbol{\mathcal{J}} \, \textbf{H}  )\, ( \textbf{A} - \textbf{I}_p ) \, \boldsymbol{\beta}  - \boldsymbol{\mathcal{J}} \, \boldsymbol{\vartheta} \bigg \rbrace \, E[ \boldsymbol{\kappa}_4 ] \Psi_{p_2 + 2} ( \chi^2_{p_2, \alpha} ; \Delta^{*} ) + E [ \boldsymbol{\kappa}_4 \, \boldsymbol{\kappa}_4 ^\prime \, I ( \chi^2_{p_2} ( \Delta^{*} ) \leq  \chi^2_{p_2 , \alpha} ) ] 
\nonumber
\end{align*}

Therefore,
\begin{align*}
\mathcal{V} ( \spe )  & =  \mathcal{V} ( \ur )  \nonumber \\
& - 2 \delta \, \bigg ( 
 \lbrace ( \textbf{I}_p - \boldsymbol{\mathcal{J}} \, \textbf{H}  )\, ( \textbf{A} - \textbf{I}_p ) \, \boldsymbol{\beta}  - \boldsymbol{\mathcal{J}} \, \boldsymbol{\vartheta}  \rbrace \, E[ \boldsymbol{\kappa}_4 ] \Psi_{p_2 + 2} ( \chi^2_{p_2, \alpha} ; \Delta^{*} ) 
\bigg ) 
\nonumber \\
& + E [ \boldsymbol{\kappa}_4 \, \boldsymbol{\kappa}_4 ^\prime \, I ( \chi^2_{p_2} ( \Delta^{*} ) \leq  \chi^2_{p_2 , \alpha} ) ]  + \delta^2  \, E ( \boldsymbol{\kappa}_4 \, \boldsymbol{\kappa}_4^\prime \, I ( \chi^2_{p_2} ( \Delta^{*} ) \leq  \chi^2_{p_2 , \alpha} ) ) 
\nonumber \\
= & \mathcal{V} ( \ur )
 - 2 \delta \, \bigg ( 
\lbrace ( \textbf{I}_p - \boldsymbol{\mathcal{J}} \, \textbf{H}  )\, ( \textbf{A} - \textbf{I}_p ) \, \boldsymbol{\beta}  - \boldsymbol{\mathcal{J}} \, \boldsymbol{\vartheta} \rbrace \, E[ \boldsymbol{\kappa}_4 ] \Psi_{p_2 + 2} ( \chi^2_{p_2, \alpha} ; \Delta^{*} ) 
\bigg )
\nonumber \\
& - \delta \, ( 2 - \delta ) \, E [ \boldsymbol{\kappa}_4 \, \boldsymbol{\kappa}_4 ^\prime \, I ( \chi^2_{p_2} ( \Delta^{*} ) \leq  \chi^2_{p_2 , \alpha} ) ] 
\nonumber \\
= & \mathcal{V} ( \ur )
 - 2 \delta \, \bigg ( 
\lbrace ( \textbf{I}_p - \boldsymbol{\mathcal{J}} \, \textbf{H}  )\, ( \textbf{A} - \textbf{I}_p ) \, \boldsymbol{\beta}  - \boldsymbol{\mathcal{J}} \, \boldsymbol{\vartheta} \rbrace 
\nonumber \\
& \times  [ \boldsymbol{\mathcal{J}} \, \textbf{H} \, ( \textbf{A} - \textbf{I}_p ) \, \boldsymbol{\beta}  + \boldsymbol{\mathcal{J}} \, \boldsymbol{\vartheta}  ]^\prime \, \Psi_{p_2 + 2} ( \chi^2_{p_2, \alpha} ; \Delta^{*} )  \bigg ) 
\nonumber \\
& - \delta \, ( 2 - \delta ) \, 
\bigg (  
\boldsymbol{\mathcal{J}} \, \textbf{H} \, \textbf{A} \, \boldsymbol{\mathcal{I}}^{-1} \, \textbf{A}^\prime \, \Psi_{p_2 + 2} ( \chi^2_{p_2, \alpha} ; \Delta^{*} )  + [ \boldsymbol{\mathcal{J}} \, \textbf{H} \, ( \textbf{A} - \textbf{I}_p ) \, \boldsymbol{\beta}  + \boldsymbol{\mathcal{J}} \, \boldsymbol{\vartheta} ] 
\nonumber \\
& \times [ \boldsymbol{\mathcal{J}} \, \textbf{H} \, ( \textbf{A} - \textbf{I}_p ) \, \boldsymbol{\beta}  + \boldsymbol{\mathcal{J}} \, \boldsymbol{\vartheta} ]^\prime  \Psi_{p_2 + 4} ( \chi^2_{p_2, \alpha} ; \Delta^{*} )
\bigg).
\nonumber
\end{align*}
For $ \delta = 1 $, $ \mathcal{V} ( \rpt )   $ reduces to
\begin{align*}
\mathcal{V} ( \rpt ) & =  \mathcal{V} ( \ur )
- 2 \, \bigg ( 
\bigg \lbrace ( \textbf{I}_p - \boldsymbol{\mathcal{J}} \, \textbf{H}  )\, ( \textbf{A} - \textbf{I}_p ) \, \boldsymbol{\beta}  - \boldsymbol{\mathcal{J}} \, \boldsymbol{\vartheta} \bigg \rbrace  
\nonumber \\
& \times [ \boldsymbol{\mathcal{J}} \, \textbf{H} \, ( \textbf{A} - \textbf{I}_p ) \, \boldsymbol{\beta}  + \boldsymbol{\mathcal{J}} \, \boldsymbol{\vartheta}  ]^\prime \, \Psi_{p_2 + 2} ( \chi^2_{p_2, \alpha} ; \Delta^{*} )  \bigg ) 
\nonumber \\
& -  \bigg (  
\boldsymbol{\mathcal{J}} \, \textbf{H} \, \textbf{A} \, \boldsymbol{\mathcal{I}}^{-1} \, \textbf{A}^\prime \, \Psi_{p_2 + 2} ( \chi^2_{p_2, \alpha} ; \Delta^{*} )  + [ \boldsymbol{\mathcal{J}} \, \textbf{H} \, ( \textbf{A} - \textbf{I}_p ) \, \boldsymbol{\beta}  + \boldsymbol{\mathcal{J}} \, \boldsymbol{\vartheta} ] 
\nonumber \\
 & \times [ \boldsymbol{\mathcal{J}} \, \textbf{H} \, ( \textbf{A} - \textbf{I}_p ) \, \boldsymbol{\beta}  + \boldsymbol{\mathcal{J}} \, \boldsymbol{\vartheta} ]^\prime  \Psi_{p_2 + 4} ( \chi^2_{p_2, \alpha} ; \Delta^{*} )
\bigg).
\nonumber
\end{align*}


Now we obtain $ \mathcal{V} ( \rs ) $ as follows:
\begin{align*}
\mathcal{V} ( \rs )  & =  \lim_{n \to \infty} \, E \Big (  \sqrt{n} ( \rs - \boldsymbol{\beta} )  \,    \sqrt{n} ( \rs - \boldsymbol{\beta} )^\prime   \Big )  
\nonumber \\
&  =   \lim_{n \to \infty} \, E \Big [  \sqrt{n} \Big ( \rr + \lbrace 1 - ( p_2 - 2 \rbrace T_n^{-1} ) \, ( \ur - \rr ) - \boldsymbol{\beta}  \Big ) 
\nonumber \\
&  =   \lim_{n \to \infty} \, E [ (  \boldsymbol{\kappa}^n_1  -  ( p_2 - 2 )\,T_n^{-1} \, \boldsymbol{\kappa}^n_4 ) \, (  \boldsymbol{\kappa}^n_1  -  ( p_2 - 2 )\, T_n^{-1} \, \boldsymbol{\kappa}^n_4  ){^\prime }]
\nonumber \\
&  = E [ ( \boldsymbol{\kappa}_1  -  ( p_2 - 2 )\, T_n^{-1} \,\boldsymbol{\kappa}_4 ) \, (  \boldsymbol{\kappa}_1 -  ( p_2 - 2 )\, T_n^{-1} \, \boldsymbol{\kappa}_4^\prime ]
\nonumber \\
&  = E [ ( \boldsymbol{\kappa}_1 \, \boldsymbol{\kappa}_1^\prime ] - 2 ( p_2 - 2 )\, \underbrace{ E [ \boldsymbol{\kappa}_4 \, \boldsymbol{\kappa}_1^\prime \, T_n^{-1} ] }_{e_3}+ ( p_2 - 2 )^2 \, \underbrace{ E [ \boldsymbol{\kappa}_4\, \boldsymbol{\kappa}_4^\prime \, \textbf{T}_n^{-2} }_{e_4} ] ,
\nonumber
\end{align*}

similar to $ e_1 $, we can write $ e_3 $  as follows:

\begin{align*}
e_3  & =  E [ \boldsymbol{\kappa}_4\, \boldsymbol{\kappa}_1^\prime \, T_n^{-1} ] 
\nonumber \\
&  =   E [ \boldsymbol{\kappa}_4\, \boldsymbol{\kappa}_4^\prime \, T_n^{-1} ] + \big [ ( \textbf{I}_p - \boldsymbol{\mathcal{J}} \, \textbf{H} ) \, ( \textbf{A} - \textbf{I}_p ) \,\boldsymbol{\beta} - \boldsymbol{\mathcal{J}} \, \boldsymbol{\vartheta} \big]\, E [ \boldsymbol{\kappa}_4 \, T_n^{-1}  ]
\nonumber \\
&  = Var [ \boldsymbol{\kappa}_4 ] \, E  \Big [ \frac{1}{\chi^2_{p_2 + 2}(\Delta^{*})} \Big ] + E [ \boldsymbol{\kappa}_4 ] \, E [ \boldsymbol{\kappa}_1^\prime ] \, E  \Big [ \frac{1}{\chi^2_{p_2 + 4}(\Delta^{*})} \Big ] 
\nonumber \\
&  + \big [ ( \textbf{I}_p - \boldsymbol{\mathcal{J}} \, \textbf{H} ) \, ( \textbf{A} - \textbf{I}_p ) \,\boldsymbol{\beta} - \boldsymbol{\mathcal{J}} \, \boldsymbol{\vartheta} \big]\, E [ \boldsymbol{\kappa}_4 \, T_n^{-1}  ],
\nonumber
\end{align*}
and by using Equation (\ref{equ22}), $ e_4 $ becomes
\begin{align*}
e_4  & =  E [ \boldsymbol{\kappa}_4\, \boldsymbol{\kappa}_4^\prime \, T_n^{-2} ]
\nonumber \\
&  =  Var [ \boldsymbol{\kappa}_4 ] \, E  \Big [ \frac{1}{(\chi^2_{p_2 + 2}(\Delta^{*}))^2} \Big ] + E [ \boldsymbol{\kappa}_4 ] \, E [ \boldsymbol{\kappa}_1^\prime ] \, E  \Big [ \frac{1}{(\chi^2_{p_2 + 4}(\Delta^{*}))^2} \Big ],
\nonumber
\end{align*}

Therefore,
\begin{align*}
\mathcal{V} ( \rs )   & =  \mathcal{V} ( \ur )
 \nonumber \\
& - 2 \, ( p_2 - 2 ) \, \bigg (
\big [ ( \textbf{I}_p - \boldsymbol{\mathcal{J}} \, \textbf{H} ) \, ( \textbf{A} - \textbf{I}_p ) \,\boldsymbol{\beta} - \boldsymbol{\mathcal{J}} \, \boldsymbol{\vartheta} \big]\, \big[ \boldsymbol{\mathcal{J}} \, \textbf{H} [ \textbf{A} - \textbf{I}_p ] \, \boldsymbol{\beta} + \boldsymbol{\mathcal{J}} \, \boldsymbol{\vartheta} \big ] \, E  \Big [ \frac{1}{\chi^2_{p_2 + 2}(\Delta^{*})} \Big ]
\bigg )
 \nonumber \\
& + ( p_2 - 2 ) \, ( p_2 - 4 )\, 
\boldsymbol{\mathcal{J}} \, \textbf{H} \, \textbf{A} \,  \boldsymbol{\mathcal{I}}^{-1} \, \textbf{A}^\prime \, 
\bigg (  
E  \Big [ \frac{1}{(\chi^2_{p_2 + 2}(\Delta^{*}))^2} \Big ] - E  \Big [ \frac{1}{\chi^2_{p_2 + 2}(\Delta^{*})} \Big ]
\bigg )
\nonumber \\
& + ( p_2 - 2 ) \, ( p_2 - 4 )\, 
[ \boldsymbol{\mathcal{J}} \, \textbf{H}  \, ( \textbf{A} - \textbf{I}_p ) \,\boldsymbol{\beta} + \boldsymbol{\mathcal{J}} \, \boldsymbol{\vartheta} ] \,
[ \boldsymbol{\mathcal{J}} \, \textbf{H}  \, ( \textbf{A} - \textbf{I}_p ) \,\boldsymbol{\beta} + \boldsymbol{\mathcal{J}} \, \boldsymbol{\vartheta} ]^\prime 
\nonumber \\
& \times \bigg (  
E  \Big [ \frac{1}{(\chi^2_{p_2 + 4}(\Delta^{*}))^2} \Big ] - E  \Big [ \frac{1}{\chi^2_{p_2 + 4}(\Delta^{*})} \Big ]
\bigg ).
\nonumber
\end{align*}


Finally, we can write $ \mathcal{V} ( \rps )  $ as follows:
\begin{align*}
\mathcal{V} ( \rps )   & =  \lim_{n \to \infty} \, E \Big (  \sqrt{n} ( \rps - \boldsymbol{\beta} )   \,    \sqrt{n} ( \rps - \boldsymbol{\beta} )^\prime   \Big )  
\nonumber \\
&  =   \lim_{n \to \infty} \, E \Big [  \sqrt{n}\, \Big ( 
\rs - ( 1 - (p_2 -2) \, T_n^{-1} ) \, I ( T_n <  p_2 -2  )\, ( \ur - \rr  )  - \boldsymbol{\beta}
\Big )  
\nonumber \\
&   \times \sqrt{n}\, \Big ( 
\rs - ( 1 - (p_2 -2) \, T_n^{-1} ) \, I ( T_n <  p_2 -2 )\, ( \ur - \rr  ) - \boldsymbol{\beta}
\Big )^\prime   \Big ]
\nonumber \\
&  = \mathcal{V} ( \rs )  - 2 E [ \boldsymbol{\kappa}_4\, \boldsymbol{\kappa}_3^\prime \, ( 1 - (p_2 -2) \, T_n^{-1} ) \, I ( T_n <  p_2 -2 ) ]
\nonumber \\
&  - 2 E [ \boldsymbol{\kappa}_4\, \boldsymbol{\kappa}_4^\prime \, ( 1 - (p_2 -2) \, T_n^{-1} )^2 \, I ( T_n < p_2 -2 ) ]
\nonumber \\
&  + E [ \boldsymbol{\kappa}_4\, \boldsymbol{\kappa}_4^\prime \, ( 1 - (p_2 -2) \, T_n^{-1} )^2 \, I ( T_n <  p_2 -2 ) ]
\nonumber \\
& = \mathcal{V} ( \rs )  - 2 \underbrace{  E [ \boldsymbol{\kappa}_4\, \boldsymbol{\kappa}_3^\prime \, ( 1 - (p_2 -2) \, T_n^{-1} ) \, I ( T_n < p_2 -2 ) ]  }_{e_5}
\nonumber \\
&  -  \underbrace{ E [  \boldsymbol{\kappa}_4\, \boldsymbol{\kappa}_4^\prime \, ( 1 - (p_2 -2) \, T_n^{-1} )^2 \, I ( T_n <  p_2 -2 ) ] }_{e_6} ,
\nonumber
\end{align*}

now we obtain $ e_5 $ 

\begin{align*}
e_5 & =  E [ \boldsymbol{\kappa}_4\, \boldsymbol{\kappa}_3^\prime \, ( 1 - (p_2 -2) \, T_n^{-1} ) \, I ( T_n <  p_2 -2 ) ]
\nonumber \\
&  =  E [ \boldsymbol{\kappa}_4\, E \lbrace \boldsymbol{\kappa}_3^\prime \, ( 1 - (p_2 - 2) \, T_n^{-1} ) \, \, I ( T_n <  p_2 -2 )  \, | \, \boldsymbol{\kappa}_4 \rbrace ]
\nonumber \\
&   = E [ \boldsymbol{\kappa}_4\,  \lbrace ( \textbf{I}_p -  \boldsymbol{\mathcal{J}} \, \textbf{H} ) \, ( \textbf{A} - \textbf{I}_p ) \,\boldsymbol{\beta}  - \boldsymbol{\mathcal{J}} \, \boldsymbol{\vartheta} \rbrace ^\prime \,  
\nonumber \\
&  \times ( 1 - (p_2 - 2) \, T_n^{-1} ) \, \, I ( T_n <  p_2 -2 ) ]
\nonumber \\
&  = \lbrace ( \textbf{I}_p -  \boldsymbol{\mathcal{J}} \, \textbf{H} ) \, ( \textbf{A} - \textbf{I}_p ) \,\boldsymbol{\beta}  - \boldsymbol{\mathcal{J}} \, \boldsymbol{\vartheta} \rbrace \,
E [ \boldsymbol{\kappa}_4\, ( 1 - (p_2 - 2) \, T_n^{-1} ) \, \, I ( T_n <  p_2 -2 ) ]
\nonumber \\
& =  \lbrace ( \textbf{I}_p -  \boldsymbol{\mathcal{J}} \, \textbf{H} ) \, ( \textbf{A} - \textbf{I}_p ) \,\boldsymbol{\beta}  - \boldsymbol{\mathcal{J}} \, \boldsymbol{\vartheta} \rbrace \,
E [ \boldsymbol{\kappa}_4 ] \, E  \Big [ \Big (1 - \frac{p_2 - 2}{\chi^2_{p_2 + 2}(\Delta^{*})} \Big ) \, I ( \chi^2_{p_2 + 2}(\Delta^{*}) <  p_2 - 2 )\Big ]
\nonumber \\
& =  \lbrace ( \textbf{I}_p -  \boldsymbol{\mathcal{J}} \, \textbf{H} ) \, ( \textbf{A} - \textbf{I}_p ) \,\boldsymbol{\beta}  - \boldsymbol{\mathcal{J}} \, \boldsymbol{\vartheta} \rbrace \,
 [ \boldsymbol{\mathcal{J}} \, \textbf{H}  \, ( \textbf{A} - \textbf{I}_p ) \,\boldsymbol{\beta}  - \boldsymbol{\mathcal{J}} \, \boldsymbol{\vartheta} + \boldsymbol{\mathcal{J}} \, \boldsymbol{\vartheta} ] \, 
\nonumber \\
&  \times E  \Big [ \Big (1 - \frac{p_2 - 2}{\chi^2_{p_2 + 2}(\Delta^{*})} \Big ) \, I ( \chi^2_{p_2 + 2}(\Delta^{*}) <  p_2 - 2 )\Big ]
\nonumber
\end{align*}
and based on Equation (\ref{equ22}), $ e_6 $ becomes
\begin{align*}
e_6 & =  E [ \boldsymbol{\kappa}_4\, \boldsymbol{\kappa}_4^\prime \, ( 1 - (p_2 -2) \, T_n^{-1} )^2 \, I ( T_n < p_2 -2 ) ]
\nonumber \\
&  =  Var ( \boldsymbol{\kappa}_4 )\, E\Big [ \Big( 1 - \frac{p_2 -2}{\chi^2_{p_2 + 2}(\Delta^{*})} \Big )^2 \, I ( \chi^2_{p_2 + 2}(\Delta^{*}) <  p_2 - 2 )  \Big ]
\nonumber \\
&   + E ( \boldsymbol{\kappa}_4 ) \, E ( \boldsymbol{\kappa}^\prime_4 )
 \, E\Big [ \Big( 1 - \frac{p_2 -2}{\chi^2_{p_2 + 4}(\Delta^{*})} \Big )^2 \, I ( \chi^2_{p_2 + 4}(\Delta^{*}) <  p_2 - 2 )  \Big ]
\nonumber \\
& = \boldsymbol{\mathcal{J}} \, \textbf{H}  \,  \textbf{A} \,  \boldsymbol{\mathcal{I}}^{-1}\, \textbf{A}^\prime
\, E\Big [ \Big( 1 - \frac{p_2 -2}{\chi^2_{p_2 + 2}(\Delta^{*})} \Big )^2 \, I ( \chi^2_{p_2 + 2}(\Delta^{*}) <  p_2 - 2 )  \Big ]
\nonumber \\
&  + [ \boldsymbol{\mathcal{J}} \, \textbf{H}  \, ( \textbf{A} - \textbf{I}_p ) \,\boldsymbol{\beta} + \boldsymbol{\mathcal{J}} \, \boldsymbol{\vartheta} ] \, 
[ \boldsymbol{\mathcal{J}} \, \textbf{H}  \, ( \textbf{A} - \textbf{I}_p ) \,\boldsymbol{\beta}  + \boldsymbol{\mathcal{J}} \, \boldsymbol{\vartheta} ]^\prime
\nonumber \\
&  \times E\Big [ \Big( 1 - \frac{p_2 -2}{\chi^2_{p_2 + 4}(\Delta^{*})} \Big )^2 \, I ( \chi^2_{p_2 + 4}(\Delta^{*}) <  p_2 - 2 )  \Big ]
\nonumber
\end{align*}

Therefore, $ \mathcal{V} ( \rps ) $ becomes
\begin{align*}
\mathcal{V} ( \rps )  & =  \mathcal{V} ( \rs ) 
\nonumber \\
& - 2\, \bigg (
 \lbrace ( \textbf{I}_p -  \boldsymbol{\mathcal{J}} \, \textbf{H} ) \, ( \textbf{A} - \textbf{I}_p ) \,\boldsymbol{\beta}  - \boldsymbol{\mathcal{J}} \, \boldsymbol{\vartheta} \rbrace \,
 [ \boldsymbol{\mathcal{J}} \, \textbf{H}  \, ( \textbf{A} - \textbf{I}_p ) \,\boldsymbol{\beta}  - \boldsymbol{\mathcal{J}} \, \boldsymbol{\vartheta} + \boldsymbol{\mathcal{J}} \, \boldsymbol{\vartheta} ] \, 
\nonumber \\
& \times E  \Big [ \Big (1 - \frac{p_2 - 2}{\chi^2_{p_2 + 2}(\Delta^{*})} \Big ) \, I ( \chi^2_{p_2 + 2}(\Delta^{*}) <  p_2 - 2 )\Big ]
\bigg )
 \nonumber \\
& - \bigg (
\boldsymbol{\mathcal{J}} \, \textbf{H}  \,  \textbf{A} \,  \boldsymbol{\mathcal{I}}^{-1}\, \textbf{A}^\prime
\, E\Big [ \Big( 1 - \frac{p_2 -2}{\chi^2_{p_2 + 2}(\Delta^{*})} \Big )^2 \, I ( \chi^2_{p_2 + 2}(\Delta^{*}) <  p_2 - 2 )  \Big ]
\nonumber \\
&  + [ \boldsymbol{\mathcal{J}} \, \textbf{H}  \, ( \textbf{A} - \textbf{I}_p ) \,\boldsymbol{\beta} + \boldsymbol{\mathcal{J}} \, \boldsymbol{\vartheta} ] \, 
[ \boldsymbol{\mathcal{J}} \, \textbf{H}  \, ( \textbf{A} - \textbf{I}_p ) \,\boldsymbol{\beta}  + \boldsymbol{\mathcal{J}} \, \boldsymbol{\vartheta} ]^\prime
\nonumber \\
&  \times E\Big [ \Big( 1 - \frac{p_2 -2}{\chi^2_{p_2 + 4}(\Delta^{*})} \Big )^2 \, I ( \chi^2_{p_2 + 4}(\Delta^{*}) <  p_2 - 2 )  \Big ]
\bigg ) .
\nonumber
\end{align*}

\newpage
\section*{Appendix 5.} \label{Appendix5}
\begin{table}[ht]
\centering
\caption{Selected variables by betaboost algorithm  for high dimensional body fat data}
\small
\begin{tabular}{rrr}
  \hline
 Number&Selected variables & coefficient in betaboost algorithm \\ 
  \hline
  1&Age & 0.0005 \\ 
  2&Abdo & 0.0384 \\ 
  3&Noise1 & -0.0067 \\ 
  4&Noise9 & -0.0140 \\ 
  5&Noise48 & -0.0064 \\ 
  6&Noise49 & -0.0267 \\ 
 7& Noise104 & -0.0061 \\ 
 8& Noise135 & 0.0174 \\ 
  9&Noise151 & 0.0130 \\ 
  10&Noise158 & -0.0069 \\ 
 11& Noise188 & -0.0057 \\ 
  12&Noise259 & -0.0074 \\ 
  13&Noise303 & -0.0187 \\ 
  14&Noise325 & 0.0208 \\ 
 15& Noise348 & 0.0066 \\ 
  16&Noise400 & -0.0120 \\ 
  17&Noise416 & 0.0065 \\ 
  18&Noise435 & -0.0198 \\ 
 19& Noise465 & -0.0316 \\ 
 20& Noise517 & 0.0066 \\ 
 21& Noise598 & 0.0061 \\ 
  22&Noise621 & 0.0130 \\ 
  23&Noise639 & 0.0198 \\ 
  24&Noise694 & 0.0155 \\ 
  25&Noise695 & 0.0281 \\ 
  26&Noise707 & -0.0073 \\ 
 27& Noise723 & 0.0077 \\ 
  28&Noise735 & 0.0130 \\ 
  29&Noise744 & -0.0058 \\ 
  30&Noise761 & -0.0065 \\ 
  31&Noise814 & -0.0062 \\ 
  32&Noise838 & -0.0065 \\ 
  33&Noise853 & -0.0067 \\ 
  34&Noise875 & 0.0124 \\ 
  35&Noise880 & 0.0262 \\ 
  36&Noise938 & 0.0126 \\ 
  37&Noise954 & 0.0177 \\ 
  38&Noise994 & -0.0058 \\ 
   \hline
\end{tabular}
\label{betaboostSelect}
\end{table}